\documentclass[
nofootinbib,
amsmath,
amssymb,
aps,
]{revtex4-2}

\usepackage{amsmath}
\usepackage{amssymb}
\usepackage{physics} 
\usepackage{graphicx}
\usepackage{dcolumn}
\usepackage{bm}
\usepackage{xcolor} 
\usepackage{color,enumerate,epsfig,fancyhdr,url,appendix,microtype,subcaption,latexsym,nicematrix,cancel}
\usepackage{ragged2e}

\newcommand{\Lagr}{\mathcal{L}}
\newcommand{\FT}{\mathcal{F}}

\DeclareMathOperator{\sgn}{sgn}

\usepackage{scalerel}
\usepackage[usestackEOL]{stackengine}
\def\dashint{\,\ThisStyle{\ensurestackMath{%
			\stackinset{c}{.2\LMpt}{c}{.5\LMpt}{\SavedStyle-}{\SavedStyle\phantom{\int}}}%
		\setbox0=\hbox{$\SavedStyle\int\,$}\kern-\wd0}\int}

\usepackage{tikz}
\usepackage{tikz-feynman}

\begin{document}
	
\title{Principle of Relativity and The Perturbative Quantum Gravity}

\author{Jinsu Kim}
\email{kjs098qazzz@gmail.com}
\affiliation{Department of Physics, KAIST, Daejeon 34141, Republic of Korea}

\author{Dongok Kim}
\affiliation{Department of Physics, KAIST, Daejeon 34141, Republic of Korea}

\date{\today}

\begin{abstract}
We develop a purely quantum theory based on the novel principle of relativity, termed the quantum principle of relativity, instead of directly applying the diffeomorphism invariance. We demonstrate that the essence of the principle can be extended into the quantum realm, maintaining the identical structures of active and passive transformations. By employing this principle, we show that quantum gravitational effects are naturally realized within the perturbatively valid theory, with general relativity emerging in large distances. We derive graviton propagators and provide several examples grounded in this novel framework.
\end{abstract}

\maketitle

\section{\label{sec:1} Introduction}
During the early 20-th century, two revolutionary theories arose in the physics society, which are general relativity (GR)~\cite{1915SPAW.......778E, 1915SPAW.......799E, einstein1915erklarung} and quantum mechanics (QM)~\cite{1901AnP...309..553P, 1905AnP...322..132E, 1913PMag...26....1B, 1924Natur.114...51D, 1925ZPhy...33..879H, 1926PhRv...28.1049S, 1928RSPSA.117..610D}. GR can describe the gravitational phenomena in a very large scale, while QM describe the world in a very small scale. Because there were attempts to unify two theories, one of the most successful theories was formulated in 1920s, called quantum field theory (QFT)~\cite{1927RSPSA.114..243D, 1929ZPhy...56....1H, 1930ZPhy...59..168H, 1930RSPSA.126..360D}, which is a quantum theory with special relativity~\cite{einstein1905elektrodynamik}.

However, it was found out to be extremely difficult to combine QFT and GR due to the lack of renormalizability~\cite{dewitt1967quantum, 1974AIHPA..20...69T}, and there is no standard theory up to these days. There were many efforts to solve this problem, and new theories were introduced such as loop gravity and supergravity~\cite{rovelli2000notes}, teleparallel gravity~\cite{aldrovandi2012teleparallel}, etc. But most of the theories focused on incorporating the gravity effects into QFT, not the philosophical part of GR. Only a few theories, such as loop quantum gravity, have the equivalence principle in the background by possessing (general) principle of relativity (PR), geometrically realized by the diffeomorphism invariance~\cite{dewitt1967quantum}. This approach was fruitful and it yielded new insights into quantum gravity. In this paper, however, we introduce a slightly different approach for achieving the PR, where limited attention has been given to the PR itself in quantum theories~\cite{davis1977relativity, dragan2020quantum}.

PR is one of the postulates mentioned when Einstein formulated special relativity~\cite{einstein1905elektrodynamik}. The PR asserts that no special frame of reference exists, and it allows us to set frames for arbitrary observers. In the cited paper, only the special principle of relativity is considered. However, it was later developed into the general principle of relativity for GR. On the other hand, diffeomorphism invariance, or general covariance, is the idea that coordinates do not exist a priori in nature, framed in the language of Riemannian geometry~\cite{hartle2003gravity}. In classical physics, PR and diffeomorphism invariance are almost equivalent, as setting the frame of reference means choosing the coordinate to be used, at least apparently.

The aim of this paper is to construct a purely quantum theory with PR, assuming physical reality underlying the quantum objects. We begin by distinguishing the position space in quantum theories from the Riemannian manifold, highlighting that the position is merely an eigenvalue of a position operator. We then explore the natural generalization of PR in the quantum realm by determining the correct approach to describe quantum version of active and passive transformation acting on position and coordinate. Subsequently, we identify a constraint on QFT that adheres to the quantum principle of relativity. Ultimately, we frame the theory within the path integral formalism. Our investigation yields a perturbatively valid quantum theory with PR, providing a quantitative description of quantum gravity.

The structure of this paper is as follows. In Sec.~\ref{sec:2}, we present a thought experiment about changing the frame in a quantum manner and argue that only a specifically constrained subspace of the Hilbert space is physical. The novel constraint, termed the condition for quantum relativity, is revealed to be crucial in maintaining the identical property of applying both active and passive quantum transformations. We show that this condition is not introduced as an additional postulate, but derived as a necessary consequence of applying PR to quantum reference frames, assuming only standard quantum mechanics and the equivalence of all inertial observers. Because these transformations generally map a manifold to another that is not isomorphic to the original one, we claim that underlying Riemannian manifold does not exist prior to the construction of the Hilbert space. 

The following sections extend this principle to a perturbative quantum theory of gravity. In Sec.~\ref{sec:3}, we apply the quantum principle of relativity to quantum field theory and examine its physical implications. By constructing the partition function, we show that the graviton naturally appears as the quantum fluctuation of the metric field, in contrast to the metric itself. In Sec.~\ref{sec:4}, we derive the explicit form of the graviton propagator under the assumption of a nearly flat background. Throughout these sections, we demonstrate that a consistent implementation of the principle requires the introduction of a ghost-like internal graviton field and a causal propagator. The renormalizability of the resulting theory is discussed in Sec.~\ref{sec:5}, and Sec.~\ref{sec:6} presents illustrative examples that highlight the physical consequences of the framework. We conclude with a summary in Sec.~\ref{sec:7}.

\section{\label{sec:2}Quantum principle of relativity}
\subsection{Thought Experiment on Rulers}
In the framework of GR, a local coordinate, denoted as $x$, is defined on spacetime to represent a specific point on a manifold. Given that spacetime possesses a physical reality in GR due to its interaction with matter, a point is often regarded as a physical entity that is universally shared, even though the value of the point is dependent on the local coordinate. Conversely, from the quantum perspective, the position $x$ of a particle is discerned as a measured value after applying a position operator, $\hat{X}$, showing one of the eigenvalues of $\hat{X}$. As the ``measured outcome" represents what an observer perceives post-experiment, a position in quantum theory is more of a readout than an intrinsic physical entity. 


Importantly, since the readout values can be defined arbitrarily, the construction of $\hat{X}$ depends on the measurement device employed, much like the choice of local coordinates. In this paper, we refer to the device used to measure position as the ``ruler''. A thought experiment arises when one considers treating the ruler itself as a state vector in Hilbert space, a natural proposition given that all rulers used in physical experiments are composed of ordinary matter existing in the universe.

Consider three identical bar rulers, denoted as ruler~A, ruler~B, and ruler~C, that can be used to measure the position of an object. Each ruler is marked with ticks to indicate position, and the spacings between these ticks are consistent. Within the framework of a Hilbert space, a single tick from ruler~A can be selected and represented as a state vector as follows:
\begin{equation}
|\psi_A\rangle = |x\rangle_A
\end{equation}
Here, $x$ denotes the position value of the tick, while $|\cdots \rangle_A$ indicates that the reading is taken from ruler~A. It's crucial to note that $|\psi_A \rangle$ must equal $|x\rangle_A$, since ruler~A is interpreting its own tick. Now, let's consider a below quantum state of the rulers.
\begin{equation}
	\label{eq:rulerA}
	|\psi_A, \psi_B, \psi_C\rangle = |x\rangle_A (|y_1\rangle|z_1\rangle + |y_2\rangle|z_2\rangle)_A
\end{equation}

If there is no specific, representative ruler in this universe, a condition that must hold true when the PR is valid, one should be able to switch the readout reference from ruler~A to ruler~B or ruler~C, since the rulers are identical. Let's consider a case where the reference is switched to ruler~C, and examine how the states of ruler~A (the one being switched), ruler~B (the one not involved in the switching), and ruler~C (the one that is to be switched to) are affected. This switching is achieved by simply reading the values from the ticks of ruler~C. Therefore, if we have $|\psi_C\rangle = |z_3\rangle_C$ by interpreting $|\psi_C\rangle$ with ruler~C itself, it becomes possible to express the full quantum state of all rulers using ruler~C instead of using ruler~A, by keeping in mind that the spacing between the ticks are equidistant.
\begin{equation}
	\label{eq:rulerC}
	|\psi_A, \psi_B, \psi_C\rangle' = (|x+\Delta_1\rangle|y_1+\Delta_1\rangle + |x+\Delta_2\rangle|y_2+\Delta_2\rangle)_C|z_3\rangle_C
\end{equation}
We have $\Delta_1 = z_3 - z_1$ and $\Delta_2 = z_3 - z_2$. Comparing Eq.~\ref{eq:rulerA} and Eq.~\ref{eq:rulerC}, it's evident that both expressions describe the same physical situation. The sole distinction lies in the ruler being utilized, suggesting that the equations convey equivalent informational quality. 

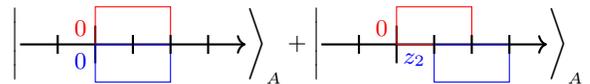
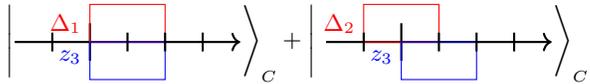
\begin{figure}
\centering
\begin{subfigure}{.5\textwidth}
\begin{equation*}
\left|
\begin{tikzpicture}[baseline={([yshift=-.5ex]current bounding box.center)}, scale=0.5]
\draw[thick, ->] (-2,-0) -- (4, 0);
\draw[thick] (0,-0.5) -- (0,0.5);
\draw[red] (0,0) node[above left]{$0$} -- (0,1) -- (2,1) -- (2,0) -- (0,0);
\draw[blue] (0,0) node[below left]{$0$} -- (0,-1) -- (2,-1) -- (2,0) -- (0,0);
\draw[thick] (-1, -0.25) -- (-1, 0.25);
\draw[thick] (1, -0.25) -- (1, 0.25);
\draw[thick] (2, -0.25) -- (2, 0.25);
\draw[thick] (3, -0.25) -- (3, 0.25);
\end{tikzpicture}
\right\rangle_A
+
\left|
\begin{tikzpicture}[baseline={([yshift=-.5ex]current bounding box.center)}, scale=0.5]
\draw[thick, ->] (-2,-0) -- (4, 0);
\draw[thick] (0,-0.5) -- (0,0.5);
\draw[red] (0,0) node[above left]{$0$} -- (0,1) -- (2,1) -- (2,0) -- (0,0);
\draw[blue] (1,0) node[below left]{$z_2$} -- (1,-1) -- (3,-1) -- (3,-0) -- (1,0);
\draw[thick] (-1, -0.25) -- (-1, 0.25);
\draw[thick] (1, -0.25) -- (1, 0.25);
\draw[thick] (2, -0.25) -- (2, 0.25);
\draw[thick] (3, -0.25) -- (3, 0.25);
\end{tikzpicture}
\right\rangle_A
\end{equation*}
\caption{Ruler~A (red) and ruler~C (blue) in frame~A}
\end{subfigure}
\begin{subfigure}{.5\textwidth}
\begin{equation*}
\left|
\begin{tikzpicture}[baseline={([yshift=-.5ex]current bounding box.center)}, scale=0.5]
\draw[thick, ->] (-2,-0) -- (4, 0);
\draw[thick] (0,-0.5) -- (0,0.5);
\draw[red] (0,0) node[above left]{$\Delta_1$} -- (0,1) -- (2,1) -- (2,0) -- (0,0);
\draw[blue] (0,0) node[below left]{$z_3$} -- (0,-1) -- (2,-1) -- (2,-0) -- (0,0);
\draw[thick] (-1, -0.25) -- (-1, 0.25);
\draw[thick] (1, -0.25) -- (1, 0.25);
\draw[thick] (2, -0.25) -- (2, 0.25);
\draw[thick] (3, -0.25) -- (3, 0.25);
\end{tikzpicture}
\right\rangle_C
+
\left|
\begin{tikzpicture}[baseline={([yshift=-.5ex]current bounding box.center)}, scale=0.5]
\draw[thick, ->] (-2,-0) -- (4, 0);
\draw[thick] (0,-0.5) -- (0,0.5);
\draw[red] (-1,0) node[above left]{$\Delta_2$} -- (-1,1) -- (1,1) -- (1,0) -- (-1,0);
\draw[blue] (0,0) node[below left]{$z_3$} -- (0,-1) -- (2,-1) -- (2,-0) -- (0,0);
\draw[thick] (-1, -0.25) -- (-1, 0.25);
\draw[thick] (1, -0.25) -- (1, 0.25);
\draw[thick] (2, -0.25) -- (2, 0.25);
\draw[thick] (3, -0.25) -- (3, 0.25);
\end{tikzpicture}
\right\rangle_C
\end{equation*}
\caption{Ruler~A (red) and ruler~C (blue) in frame~C}
\end{subfigure}
\captionsetup{justification=RaggedRight, singlelinecheck=false}
\caption{Graphical representation of Eq.~\ref{eq:rulerA} (above) and Eq.~\ref{eq:rulerC} (below), assuming $x=z_1=0$ for simplicity. Ruler~B is not included, as it is not involved in the reference switching process. Both cases represent the same physical situation; only the coordinate values differ, while the relative distances are preserved.}
\label{fig:thought}
\end{figure}

In addition to Fig.~\ref{fig:thought}, which only illustrates changes in superposition, the equations remain logically consistent even in the presence of entanglement among rulers. Suppose ruler~B and ruler~C are entangled and superposed with respect to the ticks defined by ruler~A, as represented in Eq.~\ref{eq:rulerA}. In this situation, consider a case where ruler~C measures the position of ruler~A. Since the measurement must also collapse the state of ruler~C in the frame of ruler~A to maintain the same information, and since ruler~B is entangled with ruler~C, this measurement causes the position of ruler~B to collapse accordingly. This implies that the ticks of ruler~A and ruler~B are entangled in the viewpoint of ruler~C, like described in Eq.~\ref{eq:rulerC}, and suggests that the entanglement structure may be preserved under a change of reference frame to any ruler. Based on these observations in both superposed and entangled scenarios, we propose a quantum principle of relativity.

\textit{Quantum principle of relativity:} An unified Hilbert space, denoted as $\mathcal{H}$, exists independent of the chosen ruler. A physical state is represented as a vector within this Hilbert space, while the particular ruler in use defines a position operator, not $\mathcal{H}$, by assigning appropriate eigenvalues (or readouts) to each corresponding basis state.
 
Using the aforementioned principle, one can introduce the unified Hilbert space and equate Eq.~\ref{eq:rulerA} with Eq.~\ref{eq:rulerC}, because they indicate the same vector in $\mathcal{H}$:
\begin{equation}
    \label{eq:rulerAC}
    |\psi_A, \psi_B, \psi_C\rangle = |\psi_A, \psi_B, \psi_C\rangle'
\end{equation}
For the above equation to hold true across all values of $\Delta_1$, $\Delta_2$, $x$, $y_1$, and $y_2$, we consider a new linear operator, $\hat{O}_{AC}$.
\begin{flalign}
    \hat{O}_{AC} (z_3) &= |z_3;C\rangle \int dz \langle z;C| e^{ -i \sum_{j \in {A,B}} \hat{P}_j (z_3 - z)} \\
    \label{eq:ACtrans}
    &\equiv |z_3;C\rangle \int dz \langle z;C| e^{ -i \hat{Q}_C (z_3 - z)}
\end{flalign}
Here, $\hat{P}_j$ represent the momentum operators, acting as Lie generators~\cite{1928RSPSA.117..610D}. A newly defined operator $\hat{Q}_C$ is a translator to every rulers except ruler~C. Although the subscript $C$ like $|\cdots\rangle_C$ is missing in Eq.~\ref{eq:ACtrans}, values used in $\hat{O}_{AC}$ reference ruler~C. The operator $\hat{O}_{AC}$ incorporates the value $z_3$ by the definition of ruler~C.

Implementing $\hat{O}_{AC}$, the relationship presented in Eq.~\ref{eq:rulerAC} enables one to deduce the natural transformation between rulers:
\begin{flalign}
    \label{eq:ACtrans1}
    |\psi_A, \psi_B, \psi_C\rangle &= |x\rangle_A (|y_1\rangle|z_1\rangle + |y_2\rangle|z_2\rangle)_A \\
    &= |\psi_A, \psi_B, \psi_C\rangle' \\
    \label{eq:ACtrans2}
    &= \hat{O}_{AC} |x\rangle_C (|y_1\rangle|z_1\rangle + |y_2\rangle|z_2\rangle)_C
\end{flalign}
Since the form of the ket vectors in Eq.~\ref{eq:ACtrans1} and Eq.~\ref{eq:ACtrans2} is identical, and defining $\hat{O}_{CA}$ similarly to $\hat{O}_{AC}$,
\begin{equation}
    \label{eq:OacProp}
    \therefore |\Psi\rangle_A = \hat{O}_{AC} |\Psi\rangle_C, \ |\Psi\rangle_C = \hat{O}_{CA} |\Psi\rangle_A
\end{equation}
where $\Psi$ represents the states of $(\psi_A, \psi_B, \psi_C)$ in general. 

Several properties of the operator deserve attention. Firstly, $\hat{O}_{AC}$ exhibits idempotent behavior ($\hat{O}_{AC}^2 = \hat{O}_{AC}$), which means that $\hat{O}_{AC}$ acts as a projector on $\mathcal{H}$, given that the eigenvalues of idempotent operators are restricted to being either 0 or 1. Secondly, it is shown from Eq.~\ref{eq:OacProp} that $|\Psi\rangle_A$ and $|\Psi\rangle_C$ are inter-transformable. Such behavior is feasible only if the rulers span the same subspace within $\mathcal{H}$, which is interesting as the operators are revealed as projectors. These results allow us to define a new space that is irrelevant to the ruler in use, $\mathcal{H}_\mathrm{phy}$, a subspace projected by the transformation operator. Because only the state vectors in $\mathcal{H}_\mathrm{phy}$ can be measured using position operator, we claim that $\mathcal{H}_\mathrm{phy}$ represents the physical world on behalf of $\mathcal{H}$.

Finally, but most importantly, $\hat{O}_{AC}$ absorbs right-acting $\hat{U}(a)$, where $\hat{U}(a) \equiv \exp(-ia \sum_j \hat{P}_j)$ is the total translator.
\begin{flalign}
        \hat{O}_{AC} \hat{U} 
        &= |z_3;C\rangle \int dz \langle z;C| e^{ -i \hat{Q}_C (z_3 - z + a) -i\hat{P}_C a} \\
	&= |z_3;C\rangle \int dz \langle z-a;C| e^{ -i \hat{Q}_C (z_3 -z +a)} \\
    \label{eq:OU_O}
	&= \hat{O}_{AC}
\end{flalign}
Based on our findings, we have $|\Psi\rangle_A = \hat{O}_{AC}|\Psi\rangle_C = \hat{O}_{AC} \hat{U}|\Psi\rangle_C$. This indicates that $\hat{O}_{AC}$ maps both $|\Psi\rangle_C$ and $\hat{U}|\Psi\rangle_C$ to the identical physical state. 


To investigate more deeply, let's introduce another subspace of $\mathcal{H}$, termed $\mathcal{H}_\mathrm{QR}$, where $\hat{U}|\Psi\rangle_\mathrm{QR} = |\Psi\rangle_\mathrm{QR}$ for any state $|\Psi\rangle_\mathrm{QR}\in\mathcal{H}_\mathrm{QR}$. Given the significance of the condition to define $\mathcal{H}_\mathrm{QR}$, we term it the ``condition for quantum relativity" or just the ``QR condition":
\begin{equation}
    \label{eq:QRCond}
    \hat{P}|\Psi\rangle_\mathrm{QR} \equiv \sum_j \hat{P}_j|\Psi\rangle_\mathrm{QR} = 0.
\end{equation}
Within $\mathcal{H}_\mathrm{QR}$, the operator $\hat{U}(a)$ is essentially the same as the identity operator $\hat{I}$ for any value of $a$. Furthermore, under the QR condition, $\hat{O}_{AC}$ becomes simpler.
\begin{flalign}
	\hat{O}_{AC} &= |z_3;C\rangle \int dz \langle z;C| e^{ -i \hat{Q}_C (z_3 - z)} \hat{I} \\
	&= |z_3;C\rangle \int dz \langle z;C| e^{ -i \hat{Q}_C (z_3 - z) } \hat{U}(z - z_3) \\
	&= \int dz \big( \hat{I} |z_3;C\rangle \langle z_3;C| \hat{I} \big) \\
	&= \int dz \big( \hat{U}(z) |z_3;C\rangle \langle z_3;C| \hat{U} ^\dagger (z) \big) \\
	&= \hat{I}
\end{flalign}
This implies $\mathcal{H}_\mathrm{QR} \subseteqq \mathcal{H}_\mathrm{phy}$, since it is the only way for a projector to be equivalent to the identity operator.

On the other hand, for an infinitesimally small non-zero $a$, the operator of $\hat{U}(a)$ projected onto $\mathcal{H}_\mathrm{phy}$ is given as:
\begin{flalign}
            \hat{O}_{AC} ^\dagger (z_3) (\hat{I} - i a \hat{P}) \hat{O}_{AC}(z_3) &= \hat{O}_{AC} ^\dagger (z_3) \hat{U}(a)\hat{O}_{AC}(z_3) \\
	&= \hat{O}_{AC} ^\dagger (z_3) \hat{O}_{AC}(z_3 + a) \hat{U}(a) \\
	&= \hat{O}_{AC} ^\dagger (z_3) \hat{O}_{AC}(z_3 + a)
\end{flalign}
From $\hat{O}_{AC} ^\dagger (z) \hat{O}_{AC} (z' \neq z) = 0$ according to Eq.~\ref{eq:ACtrans}, the above forbids the total translation, which makes no physical sense. The solution to this is to require $\hat{O}_{AC} ^\dagger \hat{P} \hat{O}_{AC} = 0$, thereby making $a$ effectively zero. This leads us to the opposite relationship, $\mathcal{H}_\mathrm{phy} \subseteqq \mathcal{H}_\mathrm{QR}$. 

As a result, the QR condition is both a necessary and sufficient condition for $\mathcal{H}_\mathrm{phy}$. Additionally, in the space of $\mathcal{H}_\mathrm{QR}$, not only $\hat{O}_{AC}$ but any ruler-transformation operator $\hat{O}$ becomes the identity operator, which is trivial due to the symmetries between the rulers. This offers an universal viewpoint that is independent of the specific ruler in use.

\subsection{Discussions and implications}
There are key distinctions when the quantum relativity framework is compared to traditional quantum theories. Firstly, the information of the physical coordinate (the ruler) is fully contained in the state vector. This characteristic enables one to swap the ruler that serves as the source of the coordinate, with the help of Eq.~\ref{eq:ACtrans}. Hereafter, we call this reference switching the ``quantum coordinate transformation". Since the coordinate isn't grounded on a manifold but is solely constructed by the ticks on the ruler, we also introduce the term ``quantum coordinate" to differentiate it from classical coordinates.

Secondly, a classical coordinate can be defined by the quantum coordinate, but this does not necessarily imply the existence of a classical coordinate transformation analogous to the quantum one. When the rulers are stationary and not entangled with external physical objects, they act only as manifold generators without having any physical interactions with others. In such cases, the ruler states can be taken out of the Hilbert space, apparently breaking the QR condition. Then the Hilbert space converges to the traditional one, and the quantum coordinate aligns with the classical coordinate by defining the manifold instead of explicitly expressing the ruler in use.

However, some transformations, such as the one exemplified in Eq.~\ref{eq:ACtrans2}, defy classical representation. As seen in Eq.~\ref{eq:rulerA}, the tick on ruler~A occupies a definite position denoted as $x$. Conversely, that very same tick, when referenced in Eq.~\ref{eq:rulerC}, is dispersed and superposed across two distinct positions. Such behavior violates the principle of one-to-one map between manifolds, and precludes the concept of a universally shared Riemannian spacetime in the first place, challenging the diffeomorphism invariance as a fundamental principle.

Thirdly, the total momentum of the system must vanish, ensuring the ability to transform the quantum coordinate from one to another without having logical inconsistencies. As a consequence, if one knows about the translation operator of the ruler in use, it essentially provides an ability to use the (ruler-in-use-excluded) system's total translator. This insight tells us that the transformation of the quantum coordinate encompasses not just the rulers being swapped but all particles in the system. Given that the frame of a quantum system depends on a ruler that parametrizes the position values, this characteristic can be perceived as an effect of a passive transformation. Viewing it from this angle, the QR condition is understood as a quantum counterpart of the active-passive transformation, wherein acting in both directions yields no difference.

This condition is not merely a choice of a convenient center-of-mass frame. Since momentum acts as the generator of translations, the total translator necessarily shifts both the system and the ruler, which jointly constitute $\mathcal{H}$. Because the ruler is a physical quantum object rather than a fixed classical reference frame, any non-vanishing total momentum $\hat{P} \neq 0$ induces asymmetric translations between the system and the ruler, rendering relational observables frame dependent. Requiring that relational observables remain invariant under total translations leads to the QR condition, in agreement with the logic of studies on quantum reference frames~\cite{rovelli1996relational,giacomini2019quantum,vanrietvelde2020change}.

The final distinguishing feature is the ability, through the QR condition, for a particular particle to be in rest to a position eigenstate, $|x_0\rangle$, by manipulating $\hat{U}$ distinctly for each component in the superposed state, $|x\rangle$.
\begin{flalign}
    |\Psi_\mathrm{particles} \rangle_\mathrm{QR} &= \int dx A(x) \big(|x \rangle |\Psi_\mathrm{else} (x)\rangle \big) \\
    &= \int dx A(x) \hat{U}(x_0-x) \big(|x\rangle |\Psi_\mathrm{else}(x)\rangle \big) \\
    &= |x_0\rangle |\Psi_\mathrm{else} (x_0)\rangle \cdot \int dx A(x)
\end{flalign}
$A(x)$ is a probability amplitude at position $x$. Using the above quantum coordinate, the notions of superposition and entanglement appear inapplicable to the state of the left particle, allowing one to treat it as a classical object. We coin the term ``classical frame" to describe a collection of such particles. A typical example is the ruler that is in use, which lies in the classical frame by its definition.\footnote{Just because particles within a classical frame are in position eigenstates, it doesn’t mean they remain stationary. Consider a particle with a state defined as $|x_0 + v t\rangle$, where $v$ is a relative velocity in the frame.}

The existence of the classical frame resolves the fundamental question of how superposed objects perceive their own state: they can also sense themselves as existing at a single point, because such a frame of reference is always permissible to establish. In addition, this new perspective to classical frames explains why measurement devices can be regarded as classical in quantum theories: we are used to employ our devices that align with a classical frame.

\section{\label{sec:3}Quantum relativity in QFT}
\subsection{Canonical quantization}
In this section we apply the QR condition to QFT with the requisite quantization. From the Noether's theorem~\cite{1918NGotM......235N}, the stress--energy(--momentum) tensor in Minkowski spacetime is represented by (we're using the $(-,+,+,+)$ metric sign convention):
\begin{equation}
	\label{eq:T}
	(T_N)^\mu _\nu = \sum_n (T^{(n)}_N)^{\mu} _\nu = \delta^\mu _\nu \Lagr - \sum_n {\partial \Lagr \over \partial (\partial_\mu \phi_n)} \partial_\nu \phi_n 
\end{equation}
Here, $n$ symbolizes the $n$-th quantum field. If all fields undergo canonical quantization~\cite{1925ZPhy...33..879H}, the below relation is followed.
\begin{equation}
	\label{eq:canonical}
	\left[\phi_m(x), \eta_{00} {\partial \Lagr(y^0 = x^0, \vec{y}) \over \partial (\partial_0 \phi_n)} \right\} = i \delta_{mn} \delta (\vec{x} - \vec{y})
\end{equation}
where $\eta_{\mu \nu}$ is the Minkowski metric tensor. $[\cdot, \cdot\}$ notation indicates super-commutator, which becomes anti-commutator when the both terms in the bracket are fermionic, otherwise it is just a commutator. The above equation holds true only when $y^0 = x^0$, given that quantum operators are pinpointed on a specific time-slice. Then the total momentum operator satisfies the relation below.
\begin{flalign}
        [P_{\nu \neq 0} (y^0), \phi_m (y)\} &= \left[ -\sum_n \int d^3 \vec{x} {\partial \Lagr(y^0, \vec{x}) \over \partial (\partial_0 \phi_n)} \partial_\nu \phi_n, \phi_m (y)\right\} \\
	&= \int d^3 \vec{x} \sum_n [\Pi_n(y^0, \vec{x}) \partial_\nu \phi_n(y^0, \vec{x}) , \phi_m (y)\} \\
	&= \int d^3 \vec{x} \sum_n [\Pi_n(y^0, \vec{x}), \phi_m (y)\} \partial_\nu \phi_n(y^0, \vec{x}) \\
	\label{eq:quantization}
	&= -i \partial_\nu \phi_m(y)
\end{flalign}
$\Pi_n$ is the canonical momentum of $\phi_n$. While $P_\nu (y^0)$ is originally defined through the spatial integration of $(T_N)^0 _{\nu}$, it is also shown to act as a generator of the total translation, according to Eq.~\ref{eq:quantization}. In situations where $\nu=0$, however, the uncertainty principle leads to a non-commutativity between $\partial_{\nu} \phi_m$ and $\phi_m$, which undermines the above derivation. Fortunately, Schr\"odinger's equation~\cite{1926PhRv...28.1049S} compensates for this, ensuring that the equation remains valid even for $\nu = 0$.
\begin{flalign}
		\left[H(y^0), \phi_m (y) \right\} &= \int d^3 \vec{x} [(T_N)^{00} (y^0, \vec{x}), \phi_m (y) \} \\
		&= [E(y^0), \phi_m (y) \} = i \partial_0 \phi_m (y) 
\end{flalign}
The energy operator, $E$, is also time-dependent to mark the chosen time-slice.

Merging the above results, it is clear that $\exp(-iP_\nu a^\nu)$ functions as the total translator. With this background, the QR condition within the context of QFT can be expressed as:
\begin{equation}
	\label{eq:Pcond0}
	P_\nu (x^0) = \int (T_N)_\nu ^0 (x^0, \vec{x}) d^3 \vec{x} = 0
\end{equation}
The problem is that imposing $P_\nu=0$ for all components leads to a trivial scenario: with a positive-definite Hamiltonian, the only solution is the vacuum. Here, inspired by $\delta \mathcal{S}_\mathrm{QFT} / \delta g_{\mu \nu} = T^{\mu \nu}/2$ in Minkowskian QFT (where $\mathcal{S}_\mathrm{QFT}$ is an action and $T^{\mu \nu}$ is the Belinfante–Rosenfeld tensor~\cite{belinfante1940current, belinfante1939spin, rosenfeld1940tenseur}), while $\delta \mathcal{S}_\mathrm{GR} / \delta g_{\mu \nu} = 0$ in general relativity (where $\mathcal{S}_\mathrm{GR}$ is a sum of matter action and Einstein–Hilbert action~\cite{feynman2018feynman, 1915SPAW.......844E, Hilbert1915}), we overcome this problem by promoting a metric $g_{\mu \nu}$ to a bosonic quantum field, and including it in Eq.~\ref{eq:T} to add a negative energy part to the system. The remaining issue is that identifying $(T_N)^{\mu \nu}$ with $T^{\mu \nu}$ for fermions requires an additional condition for the quantization, or a generalization of $g_{\mu \nu}$, such as the introduction of the vierbein~\cite{mcvittie1932dirac, de1962representations, collas2019dirac}. However, we do not delve into this problem in this paper. Instead, we heuristically force $(T_N)^{\mu \nu} \rightarrow T^{\mu \nu}$ in Eq.~\ref{eq:quantization} and Eq.~\ref{eq:Pcond0} for the present work.

For theories with dynamic metrics, assuming a Minkowski spacetime is no longer trivial. To effectively manage the theory, the conditions need to be generalized to tensor forms.
\begin{equation}
    \label{eq:Pcond1}
    P_\nu (x^0) = 0 \rightarrow P_\nu ^{\mu} (x^\perp) \equiv \int_V T_\nu ^{\mu} (x^\perp, \vec{x}) d^3 \vec{x} = 0
\end{equation}
$x^\perp$ denotes a parameter labeling a point along a time-like curve, $\gamma$. Since there is no preferred observer, $x^\perp$ may be specified at an arbitrary point on any $\gamma$. For each such point, $V$ denotes a space-like integrable hypersurface that is orthogonal to $\gamma$ at $x^\perp$ and extends to spatial infinity. To naturally generalize Eq.~\ref{eq:quantization} to a tensor form, the following condition should be met for each component of $P_\nu ^{\mu}$.
\begin{equation}
    \label{eq:Pcond2}
    [P_\nu ^{(n)\mu} (x^\perp), \phi_n (x)\} = -i (\vec{e}^{\, \mu} \cdot \vec{E}_\perp) \nabla_\nu \phi_n (x)
\end{equation}
Here, $\nabla_\nu$ denotes the covariant derivative defined using the Levi-Civita connection, while $\vec{e}^{\, \mu}$ and $\vec{E}_{\mu}$ are the contravariant and covariant basis vectors, respectively. $n$ is used to distinguish the quantum fields (the same $n$ used in $T^{(n)} _N$). 

By using the metric compatibility condition, $\nabla_\nu g_{\alpha \beta} = 0$, a particular case of Eq.~\ref{eq:Pcond2} for $n=g$ (the metric) is expressed as below.
\begin{equation}
    \label{eq:Pg}
    [P_\nu ^{(g)\mu} (x^\perp), g_{\alpha \beta} (x)] = 0
\end{equation}
Like the other relations involving the commutator in QFT, the locality is introduced for $g_{\alpha \beta}$ and $T^{(g) \mu} _\nu$, further refining the condition of Eq.~\ref{eq:Pg}.
\begin{equation}
    \label{eq:Tg}
    [T_\nu ^{(g)\mu} (x), g_{\alpha \beta} (y)] = 0
\end{equation}
Using $P^{(g)\mu} _\nu = - \sum_{n \neq g} P^{(n) \mu} _\nu$ from the QR condition, and assuming an additional locality with the other fields:
\begin{equation}
    \label{eq:Tg2}
    \bigg[\sum_{n  \neq g} T_{\nu} ^{(n)\mu} (x), \, g_{\alpha \beta} (y)\bigg] = 0
\end{equation}

Combining Eq.~\ref{eq:Tg2} with Eq.~\ref{eq:Tg}, one arrives at a result that the total stress--energy tensor, $T_\nu ^\mu$, must commute with $g_{\alpha \beta}$.
\begin{equation}
    \therefore [T_\nu ^{\mu} (x), g_{\alpha \beta} (y)] = 0
\end{equation}
Given that both $T_{\mu \nu} = T_\nu ^\alpha g_{\alpha \mu}$ and $g_{\mu \nu}$ are rank 2 symmetric tensors and the operators commute, the eigenstates of the operators precisely align. This opens up a new possibility of configuring the fundamental field instead of $g_{\alpha \beta}$, which could be $T_{\mu \nu}$, or even $T^\mu _\nu$.

\subsection{Path integral formalism}
By employing the path integral formalism~\cite{maggiore2005modern, feynman1948space, feynman2018theory, feynman1950mathematical}, the QR condition can be straightforwardly introduced in the theory. To naturally utilize a projector $|P^\mu _\nu = 0\rangle\langle P^\mu _\nu = 0| = \int DT^\mu _\nu \delta (P^\mu _\nu)$ to the partition function, we transfer all degrees of freedom from $g_{\alpha \beta}$ to $T^\mu _\nu$ and consider it as a fundamental field. Consequently, the QR condition transforms into a constraint for the new field, $T^\mu _\nu$, and takes the form of the partition function like below.
\begin{flalign}
    Z[J] &= \int  D\phi_n  D T^\mu _\nu \Big( \prod_j \delta \big(P^\mu _\nu \big)_j \Big) e^{i \mathcal{S}} \\
    \label{eq:ZJ0}
    &=\int  D\phi_n  D T^\mu _\nu \Big( \prod_j \delta \big( \int_j T^\mu _\nu d^3 \vec{x} \big) \Big) e^{i \mathcal{S}}
\end{flalign}
$j$ is a time-slice index, and the source term is included in $\mathcal{S}$. Let's change the field variable back to $g_{\alpha \beta}$ from $T^\mu _\nu$, since $g_{\alpha \beta}$ is more intuitive to describe a frame.
\begin{equation}
	Z[J] = \int  D\phi_n D g_{\alpha \beta} \Big( \prod_j \delta \big( P^\mu _\nu)_j \Big) \det \Big| {\partial T^\mu _\nu \over \partial g_{\alpha \beta}} \Big| e^{i \mathcal{S}}
\end{equation}

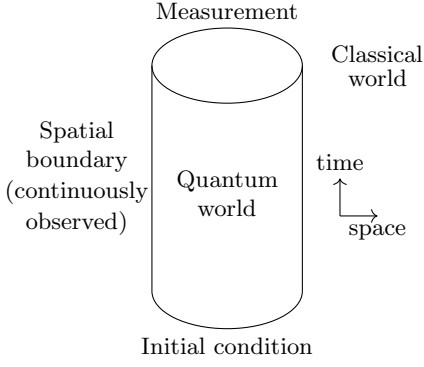
\begin{figure}
\centering
\hspace*{-1.5cm}
\begin{tikzpicture}
\draw (-1, -1.5) -- (-1, 1.5);
\draw (1, -1.5) -- (1, 1.5);
\draw (0, 1.5) ellipse (1 and 0.5);
\draw (-1, -1.5) arc (180:360:1 and 0.5);
\draw (0,-0.2) node{\shortstack{Quantum\\world}};
\draw (0,-2) node[below] {Initial condition};
\draw (0,2) node[above] {Measurement};
\draw (-2.2, 0) node{\shortstack{Spatial\\boundary\\(continuously\\observed)}};
\draw[->] (1.5,-0.5) -- (2,-0.5) node[below]{space};
\draw[->] (1.5,-0.5) -- (1.5, 0) node[above]{time};
\draw (2, 1.5) node{\shortstack{Classical\\world}};
\end{tikzpicture}
\captionsetup{justification=RaggedRight, singlelinecheck=false}
\caption{Diagram of the quantum experiment in spacetime. The quantum system is assumed to be well-regulated within the experimental setup, ensuring a finite volume despite its vast size relative to the quantum scale. The system's spatial boundary, where the measurement devices are located, is constantly monitored to observe particles escaping from it. The experiment begins by identifying the initial conditions of the fields, and ends with the measurement of the fields across the space.}
\label{fig:setup}
\end{figure}

We limit our discussion to cases when the quantum system has a finite hypervolume: the system has well-known initial condition at the temporal boundary, and the spatial boundaries are fully analyzed at each time-slice. We further limit our discussion that the objects outside the system boundaries are entirely within a classical frame by a properly defined ruler. These conditions, which actually correspond to typical quantum experimental setups (see Fig.~\ref{fig:setup}), constrain $g_{\alpha \beta}$ to fit with the boundary values. Here, we prepare various partition functions with differing metrics, to handle the cases when the allowed $g_{\alpha \beta}$ is not unique.
\begin{equation}
	\label{eq:Zg}
	Z_g[J] = \int \Big( \prod_{n} D\phi_n \Big) \det \Big| {1 \over 2} {\partial T^{\mu \nu} \over \partial g_{\alpha \beta}} \Big|  e^{ i \mathcal{S}}
\end{equation}
For later convenience, $1/2$ is inserted, and the index of $T^\mu _\nu$ is raised. By introducing a density function, $\rho(g_{\mu \nu})$, as a mediator, we now have a full partition function that can describe metric-superposed quantum systems.
\begin{equation}
	\label{eq:ZJ}
    Z = \int Dg_{\mu \nu} \, \rho (g_{\mu \nu}) Z_g \prod_j \delta \big( \int_j (\tilde{T}^\mu _\nu + T^{(g)\mu} _\nu )d^3 \vec{x} \big)
\end{equation}
$\tilde{T}^\mu _\nu$ represents the total stress--energy tensor excluding the contribution from the metric field. Notably, by identifying $T^{(g)\mu} _\nu$ with the one derived from the Einstein--Hilbert action, the metric's response to matter matches with the behavior of GR in the spatially integrated form.

By adopting the Einstein--Hilbert action, which is justified in Appendix~\ref{sec:EHLagr}, and considering that various metrics can be superposed under the control of $\rho(g_{\mu \nu})$, it is found that each metric behaves more as a semi-classical object. Yet, quantum fluctuations of spacetime still exist due to the determinant in Eq.~\ref{eq:Zg}, that can be evaluated by introducing ghost-like internal fermionic spin-2 field, $\lambda_{\alpha \beta}$.
\begin{equation}
	\label{eq:detT}
	\det \Big|{1 \over 2} {\partial T^{\mu \nu} \over \partial g_{\alpha \beta}}\Big| = \int D\lambda D \bar{\lambda} e^{-i \int d^4 x \bar{\lambda}_{\mu \nu} \big({\delta^2 \mathcal{S}[J=0] \over \delta g_{\alpha \beta} \delta g_{\mu \nu}}\big)  \lambda_{\alpha \beta} }
\end{equation}

Similar to the metric field, which is not a particle but a background entity defining trajectories, $\lambda_{\alpha\beta}$ in this theory does not exist independently on-shell. Instead, it serves as a ghost field that modifies particle paths in Feynman diagrams. In this context, we refer to this new field as the ``graviton," in contrast to the metric itself. Regarding the adequacy of a spin-2 graviton being a fermion, the spin--statistics theorem applies only to physical states in a positive-definite Hilbert space and therefore does not constrain unphysical ghost fields~\cite{weinberg1995quantum}. For this reason, the gravitons do not violate the spin--statistics relation, consistent with the standard treatment of Faddeev--Popov ghosts~\cite{faddeev1967feynman}.

Although there appear to be 16 degrees of freedom for $\lambda_{\alpha \beta}$, the true degrees of freedom are reduced to 10 due to the symmetric property of $g_{\alpha \beta}$. To address this constraint, we identify $\lambda_{\alpha \beta}$ with $\lambda_{\beta \alpha}$. Meanwhile, the free Lagrangian for the graviton is expressed as follows.
\begin{flalign}
	\label{eq:LgravStart}
	-\Lagr_{\lambda} &= \bar{\lambda}_{\mu \nu} \big({\delta^2 \mathcal{S}_\mathrm{EH} \over \delta g_{\alpha \beta} \delta g_{\mu \nu}}\big)  \lambda_{\alpha \beta} \\
    &= \bar{\lambda}_{\mu \nu} \frac{\delta}{\delta g_{\alpha \beta}} \Big( {\sqrt{-g} \over 2\kappa} \big( G^{\mu \nu} + g^{\mu \nu} \Lambda \big) \Big)  \lambda_{\alpha \beta},
\end{flalign}
where $g$ is the determinant of $g_{\mu \nu}$, $\kappa = 8\pi G$ is the Einstein gravitational constant, $G^{\mu \nu} = R^{\mu \nu} - g^{\mu \nu} R/2$ is the Einstein tensor, $R^{\mu \nu}$ is the Ricci tensor, and $\Lambda$ is the cosmological constant. By rearranging it as described in Appendix~\ref{sec:graviton} to make a shape similar to usual quantum fields, the following formula is derived:
\begin{equation}
	\label{eq:Lgrav}
        4\kappa {\Lagr_{\lambda} \over \sqrt{-g}} = 4K_\lambda - \big(R - 2\Lambda\big) (\bar{\lambda}_{\mu \nu} \lambda^{\mu \nu} - {1 \over 2} \bar{\lambda} \lambda)
        - R^{\mu \nu} (\bar{\lambda} \lambda_{\mu \nu}
        + \bar{\lambda}_{\mu \nu} \lambda - 3\bar{\lambda}_\mu ^\gamma \lambda_{\gamma \nu}) - R^\nu _{\ \alpha \mu \gamma} \bar{\lambda}^{\alpha \mu} \lambda^\gamma _\nu
\end{equation}
where,
\begin{flalign}
	K_\lambda &= -{1 \over 4} \nabla^{[\gamma} \bar{\lambda}^{\mu] \nu} \nabla_{[\gamma} \lambda_{\nu] \mu} + {1 \over 4} \nabla^{[\gamma} \bar{\lambda}^{\mu]}_\mu \nabla_{[\gamma} \lambda_{\nu]}^\nu \\
    \label{eq:K}
    &= -{1 \over 4} g_{\alpha \mu} \nabla^{[\gamma} \bar{\lambda}^{\mu][\nu} \nabla_{[\gamma} ^{} \lambda_{\nu]} ^{\alpha]}
\end{flalign}
is the kinetic term of the graviton field. This resulting Lagrangian is interesting because it includes standard kinetic terms of the form $(\nabla \lambda)^2$, along with a non-kinetic potential term. The coupling strengths of these terms are determined by the curvature. 

Throughout this process, $g_{\mu \nu}$ is decoupled from self-fluctuations and describes the classical structure of the spacetime. It follows Einstein’s equation even in the path integral formalism at sufficiently large scales, a condition for localizing the integral within the delta function. This establishes a connection to general relativity. On the other hand, the net quantum effects should also align with Einstein’s equation in the classical limit, since $\delta \mathcal{S} / \delta g_{\mu \nu} = T^{\mu \nu} = 0$ is itself a classical solution. The recursive nature of Eq.~\ref{eq:ZJ0} ensures that $T^{\mu \nu} \to 0$ even in the absence of the delta function.

In addition, $T^{\mu \nu} = 0$ in the classical limit guarantees that the QR condition is satisfied for a spacetime patch that is large enough to average out the non-minimal actions. Due to this, although the QR condition only requires the zero momentum of the whole space, it can be applied to segmented subspaces. Recalling the original motivation of the condition, this means that an observer can construct a complete and consistent quantum system to see the equivalence between the active and passive transformations, without knowing all the information of the universe.

At the fundamental level, relying on the QR condition, many field configurations are identified as representing the same quantum state. Subsequently, such configurations should be removed to prevent overcounting of states, which can be interpreted as the role played by the delta function. This constraint mechanism introduces the possibility of rendering the theory renormalizable, as discussed in the following sections.

\section{\label{sec:4}Graviton propagators}
\subsection{Perturbation theory with a cutoff scale}
In this section, we determine the graviton propagators for QR free scalar theory as a toy model. In this model, we introduce a free real scalar field, $\phi$, to an approximately flat spacetime ($R^\rho _{\ \sigma \mu \nu} \approx 0$). Under such conditions, it's allowed to define a local coordinate that is Minkowskian, denoted as $g_{\mu \nu} \approx \eta_{\mu \nu}$. Then the Lagrangian of the free scalar field is expressed like below.
\begin{equation}
    {\Lagr_\phi \over \sqrt{-g}} = -{1 \over 2}(\partial^\rho \phi \partial_\rho \phi + m^2 \phi^2)
\end{equation}
Due to the different sign convention, the sign of the kinetic term is inverted from the usual expression in QFT. Referring to Eqs.~\ref{eq:Zg}, \ref{eq:detT}, and \ref{eq:K}, the free terms for the scalar and  graviton fields are expressed as:
\begin{equation}
\Lagr_\mathrm{free} \approx -{1 \over 4\kappa} \eta_{\alpha \mu} \partial^{[\gamma} \bar{\lambda}^{\mu][\nu} \partial_{[\gamma} ^{} \lambda_{\nu]} ^{\alpha]} + \Lagr_\phi
\end{equation}
In the above, we used the approximation $\sqrt{-g} \approx \sqrt{-\eta}=1$. The next step is to obtain the graviton-scalar interaction terms described by Eq.~\ref{eq:detT}.
\begin{flalign}
        2{\Lagr_\mathrm{int} \over \sqrt{-g}} 
        &= -{(\bar{\lambda}_{\mu \nu} \lambda) \tilde{T}^{\mu \nu} \over 2 \sqrt{-g}} + 2(\bar{\lambda}_\mu ^\alpha \lambda^{\mu \beta}) \partial_\alpha \phi \partial_\beta \phi + {\bar{\lambda}_{\mu \nu} \lambda^{\mu \nu} \Lagr_\phi \over \sqrt{-g}} - {1 \over 2} (\bar{\lambda} \lambda^{\alpha \beta}) \partial_\alpha \phi \partial_\beta \phi \\
	\label{eq:LagrInt}
        &= \big( \bar{\lambda}_{\mu \nu} \lambda^{\mu \nu} - {1 \over 2} \bar{\lambda} \lambda \big) \Lagr_\phi - {1 \over 2}(\bar{\lambda} \lambda^{\alpha \beta} + \bar{\lambda}^{\alpha \beta} \lambda - 4 \bar{\lambda}_\mu ^\alpha \lambda^{\mu \beta})\partial_\alpha \phi \partial_\beta \phi
\end{flalign}
Again, $g_{\mu \nu} \approx \eta_{\mu \nu}$ has been applied. After redefining the graviton field as $\lambda_{\mu \nu} \rightarrow \sqrt{\kappa} \lambda_{\mu \nu}$ and imposing the condition $O(\kappa T^2) \ll O(1)$, where $T$ denotes the initial momentum scale of the introduced particles (i.e. the system temperature), the perturbation becomes valid.

Concurrently, to handle the Dirac delta term in Eq.~\ref{eq:ZJ}, a limit is set on the measurement sensitivity of the curvature. We define the precision scale of spacetime curvature with an order parameter, $\epsilon$.
\begin{equation}
	\Delta R^\rho _{\ \sigma \mu \nu}  = \epsilon
\end{equation}
To incorporate the condition $R^\rho_{\ \sigma \mu \nu} \approx 0$ into the partition function, the density function in Eq.~\ref{eq:ZJ} is now parameterized by $\epsilon$.
\begin{equation}
	\label{eq:Zapprox}
	Z \approx \int (D H^\mu _\nu ) Z_\eta \rho_\epsilon(H^\mu _\nu) \prod_j \delta \Big( \int_j T^\mu _\nu d^3 \vec{x} \Big)
\end{equation}
In the above expression, the measure is converted to $H^\mu _\nu \equiv G^\mu _{\nu} + \delta^\mu _{\nu} \Lambda$, so that only the curvature-related terms are considered. All the irrelevant terms are absorbed to $\rho_\epsilon$. Here, to have better understandings to the partition function, we introduce a new parameter that is more intuitive compared to $\epsilon$. Given the classical Lagrangian of the metric, $(R-2\Lambda)/2\kappa$, and specifying a spatial volume for the system, denoted as $V_0$, the new quantity $\mu_G \equiv V_0 \epsilon / 2$ serves as a suitable measure for assessing the smallest gravitational parameter ($\kappa p$) that can be detected by an observer.

Utilizing the relation $T^{(g)\mu} _\nu = - \sqrt{-g} H_\nu ^\mu / \kappa \approx - H_\nu ^\mu / \kappa$, let's examine the simplest case with zero $\Lambda$ and a uniform $\rho_\epsilon$: it has a rectangular shape with the width $2\epsilon$, and set to non-zero if $- \epsilon < H^\mu _{\nu} < \epsilon $. By applying $\rho_\epsilon$, one obtains the partition function for the approximately flat spacetime:
\begin{equation}
    \label{eq:renorm}
    Z[J] \approx Z_\eta [J] \Big|_{|\tilde{P}^\mu _\nu(x^\perp)| < |2 \mu_G / \kappa|}
\end{equation}
$\tilde{P}^\mu _\nu$ is referenced from Eq.~\ref{eq:Pcond1} and Eq.~\ref{eq:ZJ}. To retain the tensor property, it's clear that $2 \mu_G / \kappa$ must behave as a tensor, allowing us to re-express it as $\int M_\nu ^\mu (x^\perp) d^3 \vec{x} / V_0$, where $M_\nu ^\mu$ is the tensor version of $\mu_G$. Because $x^\perp$ originates from the definition of the time-slices, which gives the same results regardless of the choice made, we distinguish $x^\perp$ from the local time coordinate being used to avoid any misconceptions in handling $M_\nu ^\mu$.

Let's consider $M_\nu (x^\perp) \equiv M_\nu ^\mu \cdot e^\perp _\mu$ and treat it as a vector field, which does not require additional parameters since $M_\nu ^\mu$ is already a function of $x^\perp$. Here, for simplicity, we select a specific time coordinate, $x^\perp = t$ ($t > 0$), to define $2 \mu_G / \kappa = \int M_{t} \, d^3 \vec{x} / V_0 = M_{t}$. Combining this with Eq.~\ref{eq:renorm}, the ultraviolet (UV) cutoff emerges naturally as $\mu \equiv M_{t}$, constraining $|\tilde{P}_t| < \int \mu \, d^3 \vec{x}$.

Although $\mu$ and $\mu_G$ are good parameters to understand the system, the direct usage of them based on $M_{t}$ is not recommended due to the non-Lorentz-invariance of the time coordinate. An alternative is to bring into play $M_\rho$ as a fixed value for the given system, where $\rho=\sqrt{|r^2 - t^2|}$, and express $\mu$ via $M_\rho$.
\begin{equation}
	\mu = \frac{\partial \rho}{\partial t} M_\rho = {t \over \rho} M_\rho \sgn(t-r)
\end{equation}
$\sgn(t)$ is the sign function. The drawback with this expression is the ill-definition of $\mu$ at $\rho=0$. Considering that $\mu$ is being utilized as a regulator and that $\rho = 0$ is a divergence point in Feynman diagrams (which is discussed in the later sections), this poses a challenge. To resolve it, a different Lorentz-invariant coordinate, $s=r^2 - t^2$, is introduced. Leveraging this coordinate results in:
\begin{equation}
	\mu = {2 \over \kappa} \mu_G = \frac{\partial s}{\partial t} M_s = -2 t M_s
\end{equation}
The above indicates that $\mu_G$ has a direct proportionality to $t$, if $M_s$ is treated as a constant. Because $\mu_G = \mu_G(t)$ in this case, a more suitable governing parameter becomes $\phi_G \equiv -\kappa M_s = \mu_G / t$. This positive value, $\phi_G$, is Lorentz-invariant and serves as a measure of the finest resolution scale attainable for the gravitational potential.

As a last check, let's find conditions for the parameter scales. First, for $\mu$ to significantly surpass the momentum scale of the system, the condition $O(T) \ll O(\mu)$ is required. Also, given that the system resolution scale of $t$ aligns with $1/T$ from a wave-perspective view, the associated condition given to $\phi_G$ becomes,
\begin{equation}
	\label{eq:hierarchy0}
    O(\phi_G / T) \sim O(\mu_G) = O(\kappa \mu) \gg O(\kappa T)
\end{equation}

\subsection{Derivation of the graviton propagators}
This subsection is dedicated to derive the graviton propagators. The approach starts with rearranging the partition function:
\begin{flalign}
    Z_\lambda &\propto \int D \lambda D \bar{\lambda} \exp \big( i \int d^4 x (K_\lambda + \bar{J} \cdot \lambda + \bar{\lambda} \cdot J) \big) \\
    &= Z_0 \exp \big( i \int d^4 x \bar{J} \cdot \overset{\leftrightarrow}{\Delta} \cdot J \big)
\end{flalign}
so that,
\begin{equation}
    \bigg({1 \over Z_\lambda} \frac{\delta^2 Z_\lambda}{\delta J(x) \delta \bar{J}(y)} \bigg) \Big|_{J=0} = \langle \lambda(y) \bar{\lambda}(x) \rangle = i\overset{\leftrightarrow}{\Delta}(x,y)
\end{equation}
$J$ and $\bar{J}$ are the graviton source terms and $\overset{\leftrightarrow}{\Delta}$ is a graviton propagator expressed in tensor form. To find the right expression of $Z_\lambda$ and derive the propagator, the useful form of the Lagrangian is investigated. By additionally introducing the total derivatives that are unphysical, the free Lagrangian of graviton can be re-written in momentum basis:
\begin{equation}
	\Lagr_\lambda = \bar{\lambda}_{ij} A^{ijkl} \lambda_{kl}
\end{equation}
where the tensor $A^{ijkl}$ is obtained from Eq.~\ref{eq:K}.
\begin{flalign}
    A^{ijkl} &= {1 \over 4} \eta_{ab} \eta ^{i[b} p^{c]} \eta^{l[a} \eta ^{d]j} p_{[c} ^{}  \delta_{d]} ^k \\
    &= {1 \over 4} (\eta^{i[l} p^{c]} p_{[c} ^{} \delta_{d]} ^k \eta^{dj} - \eta^{i[j} p^{c]} p_{[c} ^{} \delta_{d]} ^k \eta^{dl})
\end{flalign}
The graviton self-interactions in Eq.~\ref{eq:Lgrav} vanish as the metric is flat and $\Lambda=0$.

The above equation shows that $p_i A^{ijkl} = p_j A^{ijkl} = p_k A^{ijkl} = p_l A^{ijkl} = 0$, meaning that only $p^i \lambda_{ij} (p) = p^j \lambda_{ij} (p) = 0$ terms survive in the action. Because gravitons are ghost-like particles that have no external lines, $\lambda_{ij} (p)$ domain can be reduced without losing the physical degrees of freedom, for the regions that do not affect to the action (if external lines exist, on the other hand, there is a possibility that the field resides in the outer region, depending on its initial conditions). Applying the idea, the Lagrangian is modified by introducing divergenceless $\lambda_{ij}$ condition, without altering the physical effects.
\begin{equation}
	\label{eq:propStart}
        \Lagr_\lambda (p) = {p^2 \over 4} \bar{\lambda}_{ij} \eta^{i[l} \eta^{j]k} \lambda_{kl} + \bar{J}^{(ij)}\lambda_{(ij)} + \bar{\lambda}_{(ij)} J^{(ij)} + \bar{\chi}^k P^{(ij)}_k \lambda_{(ij)} + \bar{\lambda}_{(ij)} P^{(ij)}_k \chi^k
\end{equation}
$J^{(ij)}$ are the source terms. $\chi^k$ and $P^{(ij)} _k$ are Grassmanian variables and coefficients, respectively, introduced to add Lagrange multipliers for giving constraints. A new notation $(ij)$ indicates that the $(ij)$ and $(ji)$ terms are identified and counted only a single time during the contraction, to avoid double counts of $\lambda_{ij}$ (as mentioned, $\lambda_{ij}$ is identified with $\lambda_{ji}$).
\begin{flalign}
	P^{(ij)}_k=\begin{pNiceMatrix}[first-row,first-col,code-for-first-row=\scriptstyle, code-for-first-col=\scriptstyle]
		 & (01) & (02) & (03) & (12) & (13) & (23) \\
		0 & p_1 & p_2 & p_3 & 0 & 0 & 0 \\
		1 & -p_0 & 0 & 0 & p_2 & p_3 & 0 \\
		2 & 0 & -p_0 & 0 & p_1 & 0 & p_3 \\
		3 & 0 & 0 & -p_0 & 0 & p_1 & p_2 
	\end{pNiceMatrix}
\end{flalign}
and,
\begin{flalign}
	P^{(ii)}_k=\begin{pNiceMatrix}[first-row,first-col,code-for-first-row=\scriptstyle, code-for-first-col=\scriptstyle]
		 & (00) & (11) & (22) & (33) \\
		0 & -p_0 & 0 & 0 & 0\\
		1 & 0 & p_1 & 0 & 0\\
		2 & 0 & 0 & p_2 & 0\\
		3 & 0 & 0 & 0 & p_3
	\end{pNiceMatrix}
\end{flalign}
The contractions of $P_k ^{(ij)} \lambda_{(ij)}$ yield nothing but $p^i \lambda_{ik}$. By rigorously computing Eq.~\ref{eq:propStart}, which is presented in Appendix~\ref{sec:propagator} in detail, the graviton propagators are expressed in a simpler form.
\begin{equation}
	\label{eq:prop}
	\begin{cases}
		\Delta_{(ij)(ij)} / 2 &= \eta^{ii} \eta^{jj} (p^2 - \eta^{ii} p_i ^2)(p^2 - \eta^{jj}p_j ^2) / p^6 \\[0.2cm]
		\Delta_{(ii)(jj)} ^{i \neq j} / 2 &= p_i ^2 p_j ^2 / p^6 - \eta^{ii} \eta^{jj} /p^2 + (\eta^{jj}p_i ^2 + \eta^{ii}p_j ^2)/p^4 \\[0.2cm]
		\Delta_{(ij)(ik)} ^{j \neq k} / 2 &= p_i ^2 p_j p_k / p^6 - \eta^{ii} p_j p_k / p^4 \\[0.2cm]
		\Delta_{(ij)(kk)} ^{i \neq j} / 2 &= p_k ^2 p_i p_j / p^6 + \eta^{kk} p_i p_j / p^4 \\[0.2cm]
		\Delta_{(ij)(kl)} ^{i \neq j \neq k \neq l}  / 2 &= p_0 p_1 p_2 p_3 / p^6
	\end{cases}
\end{equation}

As it can be clearly seen in Eq.~\ref{eq:LagrInt}, the QR scalar theory is non-renormalizable with the employment of standard propagators. To resolve this, we modify the boundary condition for the graviton by introducing causal propagators~\cite{jordan1928quantenelektrodynamik}. Contrary to Feynman propagators~\cite{schwartz2014quantum}, causal propagators utilize closed loop integration over the poles in the $p_0$ complex plane. These propagators can also be defined via other propagators.
\begin{equation}
	\Delta = {1 \over 2} (\Delta_\mathrm{retarded} - \Delta_\mathrm{advanced})
\end{equation}
An important property of causal propagators is that they are anti-symmetric with respect to $t$, since $\Delta_\mathrm{retarded}$ and $\Delta_\mathrm{advanced}$ satisfy the relationship: $\Delta_\mathrm{retarded}(t) = \Delta_\mathrm{advanced}(-t)$. Due to their anti-$t$ symmetry, causal propagators stand out for two main reasons. First, they satisfy $\Delta(t=0)=0$. This condition is crucial to negate the diverging 1-point self-loops by assigning zero values, since gravitons have the potential to create self-loops according to Eq.~\ref{eq:LagrInt}. Second, $\Delta$ is fully Lorentz-invariant, as both $\Delta_\mathrm{retarded}$ and $\Delta_\mathrm{advanced}$ maintain Lorentz invariance.

Unlike the Feynman propagators, which particles with external lines must retain in order to maintain the correct causal arrow, the anti–time-symmetric behavior of the causal propagators allows the arrows to be directed in reverse. In our case, since the graviton is a ghost-like internal field, these causal effects manifest only along internal lines of Feynman diagrams and therefore do not induce any backward information flow between external particles. This structure enables the consistent use of causal propagators without violating macroscopic causality.

Using the relation $\FT^{-1}\{ p_j F(p) \} = -i \partial_j f(x)$, where $\FT\{\cdot\}$ denotes the Fourier transform and $F(p) = \FT\{f(x)\}$, all position-basis propagators obtained from Eq.~\ref{eq:prop} can be generated by applying appropriate derivatives to $\FT^{-1}\{1/p^6\}$. After the calculations in Appendix~\ref{sec:propPosition}, the following result is obtained.

\begin{equation}
	\label{eq:Fp6Results}
	\FT^{-1}\{{128 \pi \over p^6}\} = -(r^2 - t^2) u(|t| - r) \sgn(t)
\end{equation}
Here, $u(x)$ is the Heaviside step function. 

Crucially, the fact that Eq.~\ref{eq:Fp6Results} is purely real leads to the corresponding causal propagators to be real, which would not be the case if the Feynman propagators were used. This implies, through the optical theorem, that the gravitons have zero cross section~\cite{cutkosky1960singularities, weinberg1995quantum}. For this reason, they cannot appear as external particles or as intermediate on-shell states in any diagram, and the negative-norm states do not contribute to transition probabilities~\cite{abe2019matrix}. Combined with the partial-wave unitarity ensured by the small interaction couplings, the gravitons preserve perturbative unitarity.

\section{\label{sec:5} Theory renormalization}
In this section we demonstrate a new way to make the theory perturbatively valid, using the QR free scalar theory. The renormalization, understood here as introducing UV cutoffs to render the theory physical under specific conditions, is achieved by two procedures.

\begin{enumerate}
	\item Classify diagrams without graviton lines.
	\item Add graviton lines to the classified diagrams.
\end{enumerate}

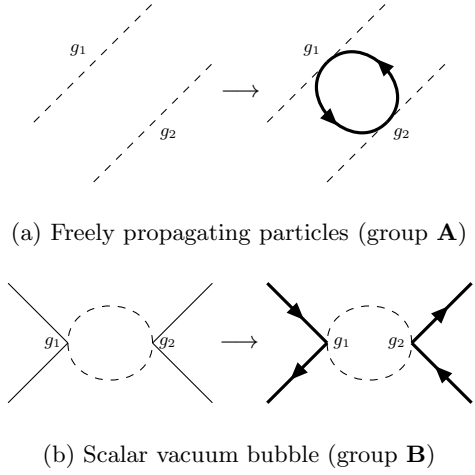
\begin{figure}
\centering
\begin{subfigure}{.45\textwidth}
\begin{equation*}
\begin{tikzpicture}[baseline={([yshift=-.5ex]current bounding box.center)}]
\begin{feynman}[scale=0.75, transform shape]
\vertex (a);
\vertex [above right=of a, label=above left:\(g_1\)] (v1);
\vertex [above right=of v1] (b);
\vertex [below right=of a] (c);
\vertex [above right=of c, label=below right:\(g_2\)] (v2);
\vertex [above right=of v2] (d);
\diagram*{
(a) -- [scalar] (b),
(c) -- [scalar] (d),
};
\end{feynman}
\end{tikzpicture}
\longrightarrow
\begin{tikzpicture}[baseline={([yshift=-.5ex]current bounding box.center)}]
\begin{feynman}[scale=0.75, transform shape]
\vertex (a);
\vertex [above right=of a, label=above left:\(g_1\)] (v1);
\vertex [above right=of v1] (b);
\vertex [below right=of a] (c);
\vertex [above right=of c, label=below right:\(g_2\)] (v2);
\vertex [above right=of v2] (d);
\diagram*{
(a) -- [scalar] (b),
(c) -- [scalar] (d),
(v1) -- [fermion, very thick, half right] (v2);
(v2) -- [fermion, very thick, half right] (v1);
};
\end{feynman}
\end{tikzpicture}
\end{equation*}
\caption{Freely propagating particles (group~\textbf{A})}
\end{subfigure}
\begin{subfigure}{.45\textwidth}
\begin{equation*}
\begin{tikzpicture}[baseline={([yshift=-.5ex]current bounding box.center)}]
\begin{feynman}[scale=0.75, transform shape]
\vertex [label=left:\(g_1\)](a);
\vertex [right=of a, label=right:\(g_2\)] (b);
\vertex [above left=of a] (v1);
\vertex [below left=of a] (v2);
\vertex [above right=of b] (v3);
\vertex [below right=of b] (v4);
\diagram*{
(a) -- [scalar, half right] (b),
(b) -- [scalar, half right] (a),
(v1) -- [transparent] (a) -- [transparent] (v2),
(v4) -- [transparent] (b) -- [transparent] (v3),
};
\end{feynman}
\end{tikzpicture}
\longrightarrow
\begin{tikzpicture}[baseline={([yshift=-.5ex]current bounding box.center)}]
\begin{feynman}[scale=0.75, transform shape]
\vertex [label=right:\(g_1\)](a);
\vertex [right=of a, label=left:\(g_2\)] (b);
\vertex [above left=of a] (v1);
\vertex [below left=of a] (v2);
\vertex [above right=of b] (v3);
\vertex [below right=of b] (v4);
\diagram*{
(a) -- [scalar, half right] (b),
(b) -- [scalar, half right] (a),
(v1) -- [fermion, very thick] (a) -- [fermion, very thick] (v2),
(v4) -- [fermion, very thick] (b) -- [fermion, very thick] (v3),
};
\end{feynman}
\end{tikzpicture}
\end{equation*}
\caption{Scalar vacuum bubble (group~\textbf{B})}
\end{subfigure}

\captionsetup{justification=RaggedRight, singlelinecheck=false}
\caption{Example diagrams for group~\textbf{A} and group~\textbf{B}, respectively. The diagrams on the left side of the arrows show the case when the graviton channels are turned off, while the right diagrams show the complete ones. Each interaction vertex has its coupling constant, $g_n$, which can be either kinetic ($\propto \kappa p_{n\alpha} p_{n\beta}$) or massive ($\propto \kappa m^2$).}
\label{fig:classifications}
\end{figure}

Diagrams derived from the QR free scalar theory, when the graviton lines are not yet drawn, can be classified based on their connectivity to external sources. The scalar particles in some diagrams, called the elements of group~\textbf{A} (see Fig.~\ref{fig:classifications}(a)), are connected to the external sources. Given the free scalar's unique feature where interaction vertices have only two scalar legs, the ends of such diagrams link to the external sources, with no intermediate branching. It's clear that these straightforward diagrams do not exhibit divergent terms.

Conversely, there are diagrams not following this structure, which are categorized to group~\textbf{B} (see Fig.~\ref{fig:classifications}(b)). In the QR free scalar theory, where tadpole type diagrams are disallowed, every one of these diagrams represents `vacuum bubbles', which are entirely closed off and not exposed externally. Unlike the traditional QFTs that usually regard vacuum bubbles as unphysical, they gain prominence in QR due to gravitons' ability to serve as a mediator between the bubble and the external environment.

The theory's structure suggests that the vacuum bubbles will manifest as loops, but their size isn't fixed. Referring to Eq.~\ref{eq:LagrInt}, each interaction vertex can be described by two types of couplings: kinetic ($\kappa \bar{\lambda} \lambda \partial \phi \partial \phi$ type) and massive ($\kappa m^2 \bar{\lambda} \lambda \phi^2$ type). If a loop has $a$ number of kinetic interaction vertices and $b$ number of mass interaction vertices, its Feynman amplitude is represented as follows:
\begin{equation}
    \label{eq:vacuumBubbles}
	\mathcal{M} \propto \int d^4 p \Big({\kappa \over p^2 + m^2} \Big) ^{a+b} \Big( \prod_{n=1} ^{a} p_{n\alpha} p_{n\beta} \Big) m^{2b}
\end{equation}
The momentum conservation is applied in the equation. $p_{n\alpha}$ and $p_{n\beta}$ refer to kinetic couplings according to the shape of $\partial_\alpha \phi \partial_\beta \phi$, which possess indices to distinguish their directivity. As it is shown in Eq.~\ref{eq:vacuumBubbles}, vacuum bubbles may induce divergences because of their loop structure. The magnitude of these divergences can be measured by introducing the parameter $\mu$ as a UV cutoff: if $a \ge 2$ and $b=0$, then $\mathcal{M} \sim O(\kappa^a \mu^4)$; if $a \ge 1$ and $b=1$, $\mathcal{M} \sim O(\kappa^{a+1} \mu^2)$; if $b = 2$, $\mathcal{M} \sim O(\kappa^{a+2} (\ln \mu))$. The diagram converge when $b > 2$.

The conventional method for treating divergences involves introducing counter terms. However, this method does not work for vacuum bubbles, as there is an infinite number of one-particle-irreducible diagrams (1PIs). In QR, on the other hand, a new approach is possible due to the concrete relation between $\kappa$ and $\mu$. By examining the strongest diverging term, where $a = 2$ and $b = 0$, and using Eq.~\ref{eq:hierarchy0}, the divergent scale is determined.
\begin{equation}
	\mathcal{M} \sim O(\kappa^2 \mu^4) = O\Big(\big( {\phi_G ^2 \over \kappa T^2} \big)^2 \Big)
\end{equation}
In the scale estimation, $t$ is omitted because it is free to set $O(t)$ as a reference scale. Due to the above, even without the traditional renormalization techniques, $\mathcal{M}$ converges if $\phi_G \ll \sqrt{\kappa} T$.

After classifying the diagrams, graviton lines must be added to generate the true diagram following the certain rules. First, all interactions in the theory have $\bar{\lambda} \lambda$ couplings, meaning that each interaction vertex in the diagram should have two graviton legs. Like the case of scalar particles in group~\textbf{A} diagrams, this implies that the chain of graviton propagators forbid branchings. Second, the chain of graviton propagators must form a closed loop. Considering that gravitons cannot form 1-point self-loops due to the use of causal propagators, and the fact that gravitons are particles without external lines, this is inevitable. Lastly, only an even number of graviton propagators are allowed in the graviton-chain loop. This constraint arises from $r$-symmetric and anti-$t$-symmetric nature of causal propagators. Graviton-chain loops exhibit fermionic directionality, and loops with opposing directions cancel out for odd-loops.

To see the renormalizability of the graviton-involved diagrams, let's separate the propagators into singular part and regular part. The Feynman propagator of the scalar field, which is another building block of the diagrams in the QR free scalar theory, is expressed in the position basis as follows.
\begin{equation}
    G(s) =
    \begin{cases}
        -{1 \over 4 \pi}\delta (s) + {m \over 8 \pi \sqrt{|s|}} H_1 ^{(1)} (m \sqrt{|s|}) & s \le 0 \\
        -{im \over 4\pi^2 \sqrt{|s|}} K_1 (m\sqrt{|s|}) & s > 0
    \end{cases}
\end{equation}
Here, $H_1 ^{(1)}$ is a Hankel function and $K_1$ is a modified Bessel function, and $s = r^2 - t^2$. As it can be seen in the equation above, the singular part of the scalar propagator exists only around $s=0$, whereas the remaining parts comprise a regular function where the integration is well-defined.

Meanwhile, an important observation from the 4-th derivatives of Eq.~\ref{eq:Fp6Results} is that all the terms in the graviton propagators contain either $\delta(|t|-r)$, $\delta'(|t|-r)$, or $\delta''(|t|-r)$. Also, given that $\delta(|t|-r)/2r = \delta(s)$, all the denominators that generate diverging poles vanish in the newly defined coordinates, $(t, s, \theta, \phi)$. In this coordinate system, the addition of graviton propagators constrains the mediated points to be light-like separated, together with additional regular effects, such as multiplication of factors or application of differentiation.

Relying on the properties of graviton propagators, the effect on the overall integrability with the scalar propagator is discussed as follows. For regular functions, the regions where integrability might fail due to the gravitons' regular effects are near boundaries. Fortunately, around the $|s|\to \infty$ boundary, which is the only boundary except $s=0$, the regular part of the scalar propagator converges to exponentially decaying continuous function in the asymptotic sense. Hence, the effects by introducing gravitons does not break the integrability of the diagrams. Since there is no way for regular functions to induce divergences except at the boundaries, the diagrams remain non-divergent upon the addition of graviton propagators.

The issue arises in the singular part of the diagram, which appears when any of the propagators in the diagram mediates points separated by a light-like distance. Fortunately, given that the singular part of the scalar propagator is also represented by a delta function, like graviton propagators, a diagram will diverge only if there are overlapping constraints set by the delta functions and their derivatives. The obvious divergences appear at 2-point loops, as they always contain terms like $\delta^{(m)}(s) \delta^{(n)}(s) \to \infty$ ($\delta^{(m)}(s)$ indicates $m$-th derivative of $\delta(s)$). In the other cases, however, no loops are divergent. To clarify this, let's consider $n$ points in 4-dimensional Minkowski spacetime (where $n > 2$), labeling them with natural numbers. Here, even if light-like separation conditions are applied between the $i$-th and $(i+1)$-th points for $i=1,2,\dots,n-1$, it does not imply that the $i=n$ point and the $i=1$ point are light-like separated. Therefore, there are no overlapping constraints for the chains of delta functions when $n > 2$.

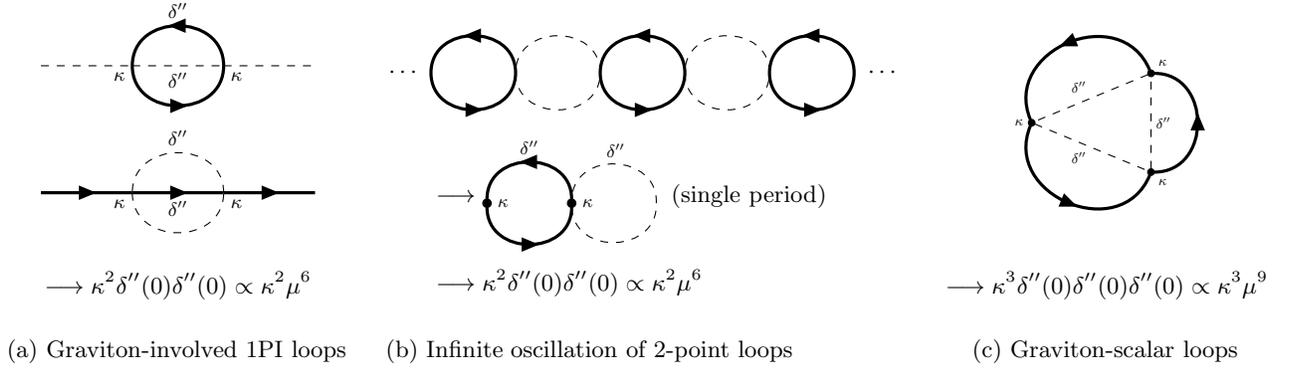
\begin{figure}[t]
\centering
\begin{subfigure}{.3\textwidth}
\begin{tikzpicture}[baseline={([yshift=-.5ex]current bounding box.center)}]
\begin{feynman}[scale=0.81, transform shape]
\vertex (a);
\vertex [right=of a, label=below left:\(\kappa\)] (b);
\vertex [right=of b, label=below right:\(\kappa\)] (c);
\vertex [right=of c] (d);
\diagram*{
(a) -- [scalar] (b) -- [scalar, edge label'=\(\delta''\)] (c) -- [scalar] (d),
(c) -- [fermion, very thick, half right, edge label'=\(\delta''\)] (b) -- [fermion, very thick, half right] (c),
};
\end{feynman}
\end{tikzpicture}

\begin{tikzpicture}[baseline={([yshift=-.5ex]current bounding box.center)}]
\begin{feynman}[scale=0.81, transform shape]
\vertex (a);
\vertex [right=of a, label=below left:\(\kappa\)] (b);
\vertex [right=of b, label=below right:\(\kappa\)] (c);
\vertex [right=of c] (d);
\diagram*{
(a) -- [fermion, very thick] (b) -- [fermion, very thick, edge label'=\(\delta''\)] (c) -- [fermion, very thick] (d),
(c) -- [scalar, half right, edge label'=\(\delta''\)] (b) -- [scalar, half right] (c),
};
\end{feynman}
\end{tikzpicture}
\begin{equation*}
\longrightarrow\kappa^2 \big(\delta''(0)\big)^2\propto\kappa^2\mu^6
\end{equation*}
\caption{Graviton-involved 1PI loops}
\end{subfigure} 
\hspace{1cm}
\vspace{0.5cm}
\begin{subfigure}{.45\textwidth}
\begin{align*}
\cdots~&
\begin{tikzpicture}[baseline={([yshift=-.5ex]current bounding box.center)}]
\begin{feynman}[scale=0.75, transform shape]
\vertex (a);
\vertex [right=of a] (b);
\vertex [right=of b] (c);
\vertex [right=of c] (d);
\vertex [right=of d] (e);
\vertex [right=of e] (f);
\diagram*{
(a) -- [fermion, very thick, half right] (b) -- [scalar, half right] (c) -- [fermion, very thick, half right] (d) -- [scalar, half right] (e) -- [fermion, very thick, half right] (f),
(f) -- [fermion, very thick, half right] (e) -- [scalar, half right] (d) -- [fermion, very thick, half right] (c) -- [scalar, half right] (b) -- [fermion, very thick, half right] (a),
};
\end{feynman}
\end{tikzpicture}
~\cdots\\
&\longrightarrow
\begin{tikzpicture}[baseline={([yshift=-.5ex]current bounding box.center)}]
\begin{feynman}[scale=0.75, transform shape]
\vertex [dot, label=right:\(\kappa\)] (a) {};
\vertex [right=of a, dot, label=right:\(\kappa\)] (b) {};
\vertex [right=of b] (c);
\diagram*{
(a) -- [fermion, very thick, half right] (b) -- [scalar, half right] (c),
(c) -- [scalar, half right, edge label'=\(\delta''\)] (b) -- [fermion, very thick, half right, edge label'=\(\delta''\)] (a),
};
\end{feynman}
\end{tikzpicture}
~\left(\mathrm{single~period}\right)\\
&\longrightarrow\kappa^2 \big(\delta''(0)\big)^2 \propto \kappa^2\mu^6
\end{align*}
\caption{Infinite oscillation of 2-point loops}
\end{subfigure} 

\begin{subfigure}{.25\textwidth}
\begin{tikzpicture}[baseline={([yshift=-.5ex]current bounding box.center)}]
\begin{feynman}[scale=0.59, transform shape]
\vertex (a);
\vertex [above left=of a, dot, label=above left:\(\kappa\)] (a2) {};
\vertex [below left=of a, dot, label=below left:\(\kappa\)] (b) {};
\vertex [below right=of a, dot, label=below right:\(\kappa\)] (c) {};
\vertex [above right=of a, dot, label=above right:\(\kappa\)] (d) {};
\diagram*{
(a2) -- [fermion, very thick, half right] (b) -- [fermion, very thick, half right] (c) -- [fermion, very thick, half right] (d) -- [fermion, very thick, half right] (a2),
(a2) -- [scalar, edge label'=\(\delta''\)] (b) -- [scalar, edge label'=\(\delta''\)] (c) -- [scalar, edge label'=\(\delta''\)] (d) -- [scalar, edge label'=\(\delta''\)] (a2),
};
\end{feynman}
\end{tikzpicture}
\begin{equation*}
\longrightarrow\kappa^4\big(\delta''(0)\big)^4\propto\kappa^4\mu^{12}
\end{equation*}
\caption{Graviton-scalar loops}
\end{subfigure}
\captionsetup{justification=RaggedRight, singlelinecheck=false}
\caption{Graviton-involved 2-point loops in QR free scalar theory. (a) Two possible 1PI loops, including graviton-graviton-scalar loop and graviton-scalar-scalar loop. Both loops have a diverging order of $O(\kappa^2 \mu^6)$. (b) Particle oscillation between gravitons and scalar particles. As the number of oscillation periods increases, the highest diverging order of a single oscillation unit also increases and eventually converges to $O(\kappa^2 \mu^6)$. (c) Graviton-scalar loops fully formed in a scalar vacuum bubble. In such cases, the highest diverging order is $O((\kappa \mu^3)^n)$, when the size of the vacuum bubble is denoted by $n$.}
\label{fig:loops}
\end{figure}

Since there are only a finite number of 2-point loops involving gravitons, renormalization is achievable. There are four types of 2-point loops with gravitons in the theory: graviton-graviton-scalar loop, graviton-scalar-scalar loop, graviton-graviton loop, and graviton-scalar loop (see Fig.~\ref{fig:loops}). For all these cases, the infinities can be represented as products of $\delta^{(n)}(0)$. Bearing in mind that the delta function arises in $\FT^{-1} \{ 1/p^2\}$ during the approximation of $\rho_\epsilon$,
\begin{equation}
	{2 \over \pi} \int_0 ^\mu dp_r \sin(p_r |t|) \sin(p_r r) \xrightarrow{\mu = \infty} \delta(|t|-r)
\end{equation}
the regulated form of the delta function can be constructed. By introducing a Lorentz-invariant cutoff $|(p_t)^2 - |\vec{p}|^2| < |M_s|$ instead of the conventional radial cutoff, $p_r < \mu$, and performing a Fourier transform, we obtain the following expression for $t > 0$:
\begin{equation}
    \label{eq:FTexact}
    {1 \over |M_s|} \FT^{-1} \{ {-4\pi \over p^2} \} = {1 \over 4\pi^2 x^2} \Big( {2K_0(x) + \pi Y_0(x)} \Big)
\end{equation}
where $x = \sqrt{|s M_s|}$. The right-hand side of Eq.~\ref{eq:FTexact} defines an integrable function that diverges logarithmically as $s \to 0$ and decays exponentially as $s \to \infty$. Since $\int_{-\infty} ^\infty \FT^{-1} \{ {-4\pi / p^2} \} ds = |M_s|$, the expression in Eq.~\ref{eq:FTexact} may be regarded as a regulated version of the delta function, $\delta(s)$, with total area preserved.

Noting that $\delta(s)$ is often approximated by a rectangular function of width $\Delta s$ and height $1/\Delta s$, we identify the corresponding width from the scale of Eq.~\ref{eq:FTexact} for estimating the divergent scale of $\delta(s=0)$.
\begin{flalign}
	\Delta s \sim {1 \over |M_s|} = {\kappa \over \phi_G}
\end{flalign}
By implementing $\Delta s$, one additionally obtains $|\delta'(s=0)| \sim 1/\Delta s^2$, and also $|\delta''(s=0)| \sim 2 / \Delta s^3$. Converting $\delta'(|t|-r=0)$ and $\delta''(|t|-r=0)$ to $\delta'(s=0)$ and $\delta''(s=0)$:
\begin{flalign}
	\frac{\partial}{\partial r} \big( 2r \delta(s) \big) &= 2 \delta(s) + 4r^2 \delta'(s) \\
	\frac{\partial}{\partial t} \big( 2r \delta(s) \big) &= -4rt \delta'(s)
\end{flalign}
and,
\begin{flalign}
	\frac{\partial^2}{\partial r^2} \big( 2r \delta(s) \big) &= 12r \delta'(s) + 8r^3 \delta''(s) \\
	\frac{\partial^2}{\partial t^2} \big( 2r \delta(s) \big) &= -4r \delta'(s) + 8r t^2 \delta''(s) \\
	-\frac{\partial}{\partial r}\frac{\partial}{\partial t} \big( 2r \delta(s) \big) &= 4t \delta'(s) + 8r^2 t \delta''(s)
\end{flalign}
One can see that $\delta'(|t|-r=0)$ have $\phi_G ^2 / \kappa^2 (\propto \mu^2)$ terms while $\delta''(|t|-r=0)$ have $\phi_G ^2 / \kappa^2$ and $\phi_G ^3 / \kappa^3 (\propto \mu^3)$ terms, showing their scales.

By using $\delta^{(n)}(0)$, it is now possible to determine the diverging order of the loops at the maximum level. For graviton-graviton-scalar loops, there are three delta functions multiplied, causing two of them to diverge. Noting that the highest diverging order of each graviton or kinetically coupled scalar propagator is $\mu^3$ (massively coupled scalar propagator has less diverging order), the leading scale will be $\kappa^2 \mu^6 \propto (\phi_G ^3 / \kappa^2)^2$. To address these terms perturbatively, the system momentum scale, $T$, must satisfy $O(\phi_G ^3 / \kappa^2) \ll O(T^4)$. Alongside the previously mentioned conditions, $O(\kappa T^2) \ll O(\phi_G)$ and $O(\phi_G) \ll O(\sqrt{\kappa} T)$, the following hierarchy is established for a valid theory:
\begin{equation}
	\label{eq:hierarchy}
    O(\kappa T^2) \ll O(\phi_G) \ll O\big((\kappa T^2)^{2/3}\big)
\end{equation}
This condition requires that the gravitational potential of a single scalar particle should be immeasurable, thereby ensuring the flatness of the metric. Simultaneously, the gravitational potential's resolution must be sufficient enough to constrain the kinetic fluctuation of virtual particles.

The same logic can be applied to graviton-scalar-scalar loops, as it is shown in Fig.~\ref{fig:loops}(a): the highest order of $\mu$ in graviton-scalar-scalar loops is also $O(\kappa^2 \mu^6) \sim O((\phi_G ^3 / \kappa^2)^2) \ll O(T^8)$. In addition, graviton-graviton loops can be addressed in the same way. The only tricky case is when each vertex of graviton-graviton loops is connected to the scalar loop, like the ones illustrated in Fig.~\ref{fig:loops}(b). The highest diverging order for such cases arises when a large chain of oscillating graviton-graviton loops and kinetic scalar-scalar loops is formed. In this case, the converging limit of the highest diverging order of a single unit is $O(\kappa^2 \mu^6)$, identical to the case of graviton-graviton-scalar loops.

Lastly, for graviton-scalar loops: if the loops form outside the scalar loops, $\mu$ can be handled straightforwardly. However, if they form within a scalar loop, complications show up as divergence from the scalar vacuum bubble is also involved, like an example of Fig.~\ref{fig:loops}(c). To solve this problem, let us consider the most divergent case, where gravitons connect all vertices of a vacuum bubble, maximizing the number of graviton-scalar loops. In such case, with a vacuum bubble size of $n$, there are $n$ number of both $\kappa$ and graviton-scalar loops. Therefore, the highest diverging order is $O((\kappa \mu^3)^n)$, just like other types of 2-point loops. As a result, all infinities, including those from vacuum bubbles, graviton-involved 2-point loops, and hybrids of the two, are effectively regulated, completing the renormalization of the theory.

Although the theory is shown to be well controlled at the perturbative level, Eq.~\ref{eq:hierarchy} may still seem challenging to be understood, particularly the right-side inequality. This complexity is simplified by recalling that $\mu_G (t)$ represents the sensitivity of the gravitational constant ($\kappa p$), and by introducing the momentum phase space from statistical mechanics.
\begin{flalign}
	{\phi_G \over \kappa T^2} &= {(\mu_G ^\mathrm{max} / \kappa) / T \over R_0 T}\\
        \label{eq:numberUncertainty}
	&= {(\Delta \sqrt{p}\,)^2 / T \over \text{(number of }1/R_0\text{\ steps in }T)} 
\end{flalign}

In the above, $\phi_G$ is re-expressed as $\kappa \cdot (\mu_G ^\mathrm{max} / \kappa) / R_0$, where $R_0$ is the system (temporal) size. In the numerator, $\mu_G ^\mathrm{max} / \kappa$ is interpreted as the uncertainty in the momentum, which is denoted as $(\Delta \sqrt{p})^2$ to achieve square-summed accumulation in the uncertainty through volume integration. By dividing $(\Delta \sqrt{p})^2$ by $T$, the domain of the uncertainty is converted to the particle number. 

Meanwhile, in the denominator, dividing $T$ by the reciprocal of the system size yields the number of minimum momentum steps within the given energy level. Then, the ratio in Eq.~\ref{eq:numberUncertainty} can be associated with $(\Delta \sqrt{N})^2$, a quantity that represents the averaged uncertainty in the particle number for each $(1/R_0)^3$ cubic volume of the momentum space.
\begin{equation}
    \label{eq:N_relation}
	{\phi_G \over \kappa T^2} \sim \big(\Delta \sqrt{N}\, \big)^{2/3}
\end{equation}
In the above, $(\Delta \sqrt{N})^2$ is raised to the one-third power to extend its domain from one to three dimensions. By using $\Delta \sqrt{N}$ instead of $\phi_G$, Eq.~\ref{eq:hierarchy} is restated as:
\begin{equation}
    \label{eq:N_uncertainty}
	O(1) \ll O\Big(\Delta \sqrt{N} \Big) \ll O\Big({T_{\text{Planck}} \over T}\Big)
\end{equation}
where $T_{\text{Planck}}$ indicates the Planck temperature.

The left-side inequality states that $\Delta \sqrt{N}$ must exceed 1. This is a standard condition in quantum experiments, as $\Delta \sqrt{N} \le 1$ implies micromanagement at the single particle level. It presents a challenge in defining the quantum system, and leads to the capability of pinpointing the particles with a precision surpassing the minimum momentum step, ($1/R_0$). The right-side inequality, on the other hand, addresses the need to prevent excessively high energy density, precluding the system from becoming gravity-dominated. When the system temperature approaches $T_{\text{Planck}}$, even individual particles acquire immense energy, undermining the assumptions of perturbation theory. Consequently, we conclude that Eq.~\ref{eq:hierarchy} emerges not merely as an acceptable condition, but as a natural description of the validity of our approximation in the perturbative quantum systems.

\section{\label{sec:6}Examples}
\subsection{Gravitational self-energy of the scalar field}
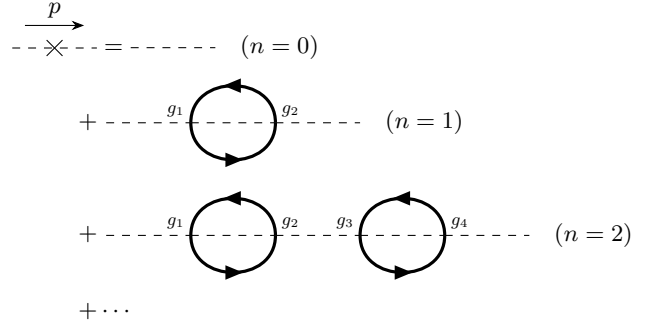
\begin{figure}
	\centering
	\begin{align*}
		\begin{tikzpicture}[baseline={([yshift=-2.7ex]current bounding box.center)}]
			\begin{feynman}[scale=0.75, transform shape]
				\vertex (a);
				\vertex [right=of a] (b);
				\diagram*{
					(a) -- [scalar, insertion=0.5, momentum=\(p\)] (b),
				};
			\end{feynman}
		\end{tikzpicture}
		&=
		\begin{tikzpicture}[baseline={([yshift=-.5ex]current bounding box.center)}]
			\begin{feynman}[scale=0.75, transform shape]
				\vertex (a);
				\vertex [right=of a] (b);
				\diagram*{
					(a) -- [scalar] (b),
				};
			\end{feynman}
		\end{tikzpicture}
		\quad(n=0)\\
		+&~
		\begin{tikzpicture}[baseline={([yshift=-.5ex]current bounding box.center)}]
			\begin{feynman}[scale=0.75, transform shape]
				\vertex (a);
				\vertex [right=of a, label=above left:\(g_1\)] (b);
				\vertex [right=of b, label=above right:\(g_2\)] (c);
				\vertex [right=of c] (d);
				\diagram*{
					(a) -- [scalar] (b) -- [scalar] (c) -- [scalar] (d),
					(b) -- [fermion, very thick, half right] (c) -- [fermion, very thick, half right] (b),
				};
			\end{feynman}
		\end{tikzpicture}
		\quad(n=1)\\
		+&~
		\begin{tikzpicture}[baseline={([yshift=-.5ex]current bounding box.center)}]
			\begin{feynman}[scale=0.75, transform shape]
				\vertex (a);
				\vertex [right=of a, label=above left:\(g_1\)] (b);
				\vertex [right=of b, label=above right:\(g_2\)] (c);
				\vertex [right=of c, label=above left:\(g_3\)] (d);
				\vertex [right=of d, label=above right:\(g_4\)] (e);
				\vertex [right=of e] (f);
				\diagram*{
					(a) -- [scalar] (b) -- [scalar] (c) -- [scalar] (d) -- [scalar] (e) -- [scalar] (f),
					(b) -- [fermion, very thick, half right] (c) -- [fermion, very thick, half right] (b),
					(d) -- [fermion, very thick, half right] (e) -- [fermion, very thick, half right] (d),
				};
			\end{feynman}
		\end{tikzpicture}
		\quad(n=2)\\
		+&\cdots
	\end{align*}
	\captionsetup{justification=RaggedRight, singlelinecheck=false}
	\caption{Infinite summation of graviton-graviton-scalar loops. The index $n$ denotes a number of involved loops. The interaction couplings, $g_n$, can be either massive or kinetic, resulting in each loop having four possible combinations: massive-massive, kinetic-kinetic, kinetic-massive, and massive-kinetic.}
	\label{fig:example1}
\end{figure}

As an initial example, we determine the gravitational self-energy of a scalar particle. The self-energy diagrams are computed by collecting all 1PIs that have single scalar lines at both the beginning and the end. It's notable that both the massive and kinetic couplings attain their parameters exclusively in 2-point graviton-graviton-scalar loops, so that an infinite summation over $n$ number of such loops will be the key, as illustrated in Fig.~\ref{fig:example1}. To approach this idea systematically, we begin by disabling the kinetic coupling terms, and compute the infinite sum of the massively coupled loops. The corresponding Feynman amplitude, parameterized by the momentum $p$, is expressed as follows.
\begin{flalign}
    \mathcal{M} &= {1 \over -p^2 - m^2} \Big( 1 - {1 \over -p^2 - m^2} \Sigma(p) + \cdots \Big) \\
    &= {1 \over -p^2 - m^2 + \Sigma(p)}
\end{flalign}
Here, $\Sigma(p)$ is a graviton-graviton-scalar single loop that can be computed in the position space.

As the graviton constrains the 2-point to be separated in light-like manner, the scalar propagator becomes $-\delta(|t|-r)/8\pi r$. Introducing a displacement between the interaction points of the loop, $w^\mu = (t,x,y,z)$, and using Eq.~\ref{eq:LagrInt},
\begin{equation}
    \label{eq:Sigma1}
        \Sigma = {\kappa^2 m^4 \over 16} \int d^4 w e^{ip\cdot w} {\delta(|t|-r) \over 8\pi r} \times \big( \Delta_{ijkl} \Delta^{ijkl} - \Delta_{ijk}^{\ \ \ k} \Delta^{ijl}_{\ \ \ l} + {1 \over 4} \Delta \Delta \big)
\end{equation}
During the computation of $\Sigma$ one faces various infinities: $\delta(0)$, $\delta'(0)$, and $\delta''(0)$. By renormalizing the infinities using Eq.~\ref{eq:FTexact}, as mentioned in the previous section, the value becomes finite after the analytical simplification process.
\begin{flalign}
    \label{eq:Sigma2}
	\Sigma &\approx {1 \over 4000} {\kappa^2 m^4 \over 4\pi^3} \big( {\phi_G \over \kappa} \big)^2 \int d^3 \vec{r} {\cos(p_0 r) e^{ip_r r \cos \theta} \over r} \\
	&= {1 \over 4000} {\kappa ^2 m^4 \over \pi^2 p_r} \big( {\phi_G \over \kappa} \big)^2 \int_0 ^\infty dr \cos(p_0 r) \sin(p_r r) \\
    &= {1 \over 4000} {\kappa ^2 m^4 \over \pi^2 p_r} \big( {\phi_G \over \kappa} \big)^2 \int_{-\infty} ^\infty dr {-i \sgn(r) \over 4} \big(e^{i(p_0 + p_r)r} - e^{i(p_0 - p_r)r}\big) \\
	&= {1 \over 4000} {\kappa^2 m^4 \over \pi^2} \big( {\phi_G \over \kappa} \big)^2  {1 \over p_r ^2 - p_0 ^2}
\end{flalign}
To obtain Eq.~\ref{eq:Sigma2} from Eq.~\ref{eq:Sigma1}, the regulated form of $f(s) = s^m \prod_i \left(\delta^{(i)}(s)\right)^{n_i}$ type functions is expanded to $\sum_m a_m \delta^{(m)}(s)$ by evaluating $a_m = \int (-s)^m f(s) ds$, based on the distributional identity: $\int (-s)^m \delta^{(n)}(s) ds = \delta_{mn}$.

Using the above results, it is possible to renormalize the scalar mass as an effect of the self-energy diagrams. The equation for $m_\mathrm{add}$ is then derived as below.
\begin{equation}
	{m_\mathrm{add} ^2 \over m^2} \approx {1 \over 4000} \big({\phi_G  \over \pi}\big)^2 \big({m^2 \over - p^2}\big) 
\end{equation}
$m_\mathrm{add}$ affects the scalar propagator, and the mass of inertia is adjusted by the given gravitational potential sensitivity, $\phi_G$. On the other hand, the gravitational mass remains to $m$.

The infinite sum of the kinetic graviton-graviton-scalar loops without the massive couplings is also computed to complete the mass renormalization.
\begin{flalign}
    \mathcal{M} &= {1 \over -p^2 - m^2} \Big(1 + h^{ij}(p) {p_i p_j \over -p^2 - m^2} + \cdots \Big) \\
    &= {1 \over - (\eta^{ij} + h^{ij}) p_i p_j - m^2}
\end{flalign}
Unlike the massively coupled case, it has plus sign in the summation, as $p_i$ and $p_j$ absorbed the imaginary numbers. To obtain the explicit form of the leading terms of $h^{ij}(p)$, below integral is computed.
\begin{flalign}
	\Sigma' &\approx {1 \over 5500} {\kappa^2 \over 4\pi^3} \big( {\phi_G \over \kappa} \big)^2 \int d^3 \vec{r} {\cos(p_0 r) e^{ip_r r \cos \theta} \over r} \\
        &= {1 \over 5500} {\kappa^2 \over \pi^2} \big( {\phi_G \over \kappa} \big)^2 {1 \over p^2}
\end{flalign}
Without loss of generality, $z$-direction is defined as a direction of $\vec{p}$, so that $h^{ij}$ can be reduced to $2\times2$ matrix. Using $\Sigma'$:
\begin{equation}
	h^{ij} \approx {1 \over 5500} {\kappa^2 \over - \pi^2 p^2} \big( {\phi_G \over \kappa} \big)^2 \begin{pmatrix}
		-p_0 ^2 & p_0 p_r \\
		p_0 p_r & -p_r ^2
	\end{pmatrix}
\end{equation}
and thus,
\begin{equation}
	h^{ij} p_i p_j \approx {1 \over 5500} \big({\phi_G \over \pi}\big)^2 p^2
\end{equation}

Absorbing the results to $m_\mathrm{add}$ again, the final form of the renormalized scalar propagator in momentum space is obtained. Because the effects from the kinetic-massive (and massive-kinetic) coupling loops vanish for real scalar fields, according to the calculation, the result becomes:
\begin{equation}
	\mathcal{M}_\mathrm{renorm.} = {1 \over - p^2 - (m^2 + m_\mathrm{add} ^2)}
\end{equation}
where,
\begin{equation}
    m_\mathrm{add} ^2 \approx \big({\phi_G \over \pi}\big)^2 \Big( {m^2 \over 4000} \big({m^2 \over -p^2}\big) - {(-p^2) \over 5500} \Big)
\end{equation}
As it is described in the equation, the mass parameter running becomes significant as $p^2$ approaches zero. For this reason, $O(|p|) \ll O(\phi_G m)$ must be met to see the mass parameter running. If the accessible resolution is worse to capture such small $|p|$, the self-energies are hidden behind the uncertainty level, and the propagator converges to the one in the traditional QFT.

\subsection{Temperature of the scalar particle with an adjustable mass}
In this subsection, the thermal temperature around a scalar particle is estimated. To begin with, the scalar particle, denoted as $\varphi_1$ in its field expression, is assumed to have an adjustable mass, $M$. In this way, this particle can be treated as a model for a Schwarzschild black hole. Then, we introduce another scalar field with zero rest mass, $\varphi_2$, and set it to have no direct interaction with $\varphi_1$ (however, the graviton interaction must be present for the system to be QR).

In this case, let us consider an event in which a stationary $\varphi_1$ spontaneously decays and emits $\varphi_2$ particles. The created $\varphi_2$ particles propagate in opposite directions, each carrying momentum, $p$. The leading contribution to this process is the two-particle pair creation, in which $\varphi_1$ loses $M$ by an amount of $2p$ ($M \gg p > 0$). As a first-order approximation, we consider only this contribution when computing the Feynman amplitude. Due to spherical symmetry, the propagation direction of $\varphi_2$ may be chosen along the $z$-axis without loss of generality. Applying the same tree-level methods used in the previous example, we obtain the following amplitude:
\begin{equation}
        \mathcal{M}_{0 \rightarrow 1} \approx \kappa^2 M^2 p^4 (M-2p) (0.009 M + 0.018 p) \big( {\phi_G \over \kappa} \big)^3 \int d^4 w (r^4 - r^2 z^2) {1 \over 2r} \delta(|t| - r) e^{2 i p t}
\end{equation}
Here, $0$ and $1$ in $\mathcal{M}_{0 \rightarrow 1}$ denote the state of $\varphi_1$ before and after the emission, respectively. 

As a second step, $\mathcal{M}_{1 \rightarrow 0}$ is computed for the case where the scalar mass, $(M-2p)$, becomes $M$ by the absorption of two $\varphi_2$ particles.
\begin{equation}
        \mathcal{M}_{1 \rightarrow 0} \approx \kappa^2 M^2 p^4 (M-2p) (0.009 M - 0.054 p) \big( {\phi_G \over \kappa} \big)^3 \int d^4 w (r^4 - r^2 z^2) {1 \over 2r} \delta(|t| - r) e^{-2 i p t}
\end{equation}
When the system is in thermal equilibrium, the transition rate from state~0 to state~1 must match the reverse transition. Denoting $n_0$ and $n_1$ as the occupation numbers for each state in the grand canonical ensemble,
\begin{equation}
    \label{eq:M_ratio1}
    {|\mathcal{M}_{0 \rightarrow 1}|^2 \over |\mathcal{M}_{1 \rightarrow 0}|^2} = {n_1 \over n_0} \simeq \exp\big({ (N_1 - N_0) \beta \mu}\big)
\end{equation}
In the above, the right-hand side is derived from the generalized Boltzmann factor, where $\beta$ indicates the inverse temperature of the system. $\mu = p$ is the chemical potential, and $N_i$ is the particle number of $\varphi_2$ particles having momentum $p \hat{z}$. Expanding both sides of the equation to first order in $p$ gives:
\begin{equation}
    \label{eq:M_ratio2}
    {|\mathcal{M}_{0 \rightarrow 1}|^2 \over |\mathcal{M}_{1 \rightarrow 0}|^2} \approx 1 + 16 {p \over M} \simeq 1 + (N_1 - N_0) \beta p
\end{equation}

For the pair-created particles, there exists a fundamental uncertainty in particle number between the states, as described in Eq.~\ref{eq:N_uncertainty}. According to this relation, $N_1 - N_0 \sim (\Delta \sqrt{N_z})^2$, with:
\begin{equation}
    \label{eq:Nz_relation}
    (\Delta \sqrt{N_z})^2 = {1 \over \sqrt{3}} (\Delta \sqrt{N})^{2/3}
\end{equation}
In the expression, the one-third power is introduced to account for the one-dimensionality of $N_z$, while the factor $1/\sqrt{3}$ arises from the angular normalization of the spherical harmonics when aligning the momentum direction to the $z$-axis. Using eqs.~\ref{eq:N_relation}, \ref{eq:M_ratio2}, and \ref{eq:Nz_relation}:
\begin{equation}
    {16 \over M} = {\beta \over \sqrt{3}} {\phi_G \over \kappa M^2} \rightarrow \beta \approx 4\pi {28 r^* \over \phi_G}
\end{equation}
$r^* = \kappa M / 4\pi$ is the Schwarzschild radius of $\varphi_1$. Finally, we introduce a physical radius, $r_\mathrm{phy} \simeq 28 r / \phi_G$, yielding the Hawking temperature, $\beta = 4\pi r^*_\text{phy}$~\cite{hawking1975particle}. Although the numerical relation between $r$ and $r_{\mathrm{phy}}$ remains uncertain and may depend on the specific model, the present treatment is sufficient to capture the expected behavior within our framework.

\section{\label{sec:7}Conclusion}
In this study, we present a novel approach to quantum gravity, which is strongly tied to the idea of integrating the principle of relativity into quantum theory. Instead of directly introducing general relativity, our theory extends the principle of relativity to the quantum domain. The application of this principle introduces a new constraint to the theory, which we term the QR condition. This condition was found essential for preserving identical structures in the application of active and passive transformations. Given that the theory is based on a newly introduced fundamental principle, the QR condition has led us to define clear concepts of both the quantum coordinate and the classical frame. As a consequence, it reveals the role of the quantum observer: an individual who defines a quantum coordinate, aligning themselves with a classical frame.

Our results suggest that the most natural way to construct a dynamic universe is to promote the metric tensor to a quantum field, with its corresponding Lagrangian being identical to the Einstein--Hilbert action. Using the path integral formalism, we applied the QR condition and derived the full expression of the partition function. Based on our study, the quantum fluctuation of the metric field can be decoupled, while the semi-classical part of the metric obeys the Einstein field equations in the integrated form. We further analyzed the quantum fluctuation components, which manifest as a ghost-like internal field that we refer to as the graviton.

By introducing a free scalar field in an approximately flat spacetime, our results demonstrate that gravitational effects can be naturally embedded in a quantum theory. Assuming graviton propagators as causal propagators under a few acceptable scale conditions, we have explicitly computed its form in position space. We have further shown that the theory is perturbatively valid, with examples highlighting the practical implications of our approach. In particular, the gravitational self-energy suggests that the running of the mass parameter may uncover gravitational quantum effects, if the system's momentum resolution reaches below $\phi_G m$. Future studies may explore applications of the theory, potentially bridging the long-standing gap between quantum mechanics and general relativity.

Although several new postulates and assumptions have been introduced, our approach provides fresh insights into the understanding of the quantum world and the renormalization strategy. There is also much more to explore in terms of conceptual development, but we leave such investigations for future work.

\acknowledgments
The authors would like to give special thanks to Dr. Junu Jeong, Dr. Danho Ahn, and Dr. Younggeun Kim for their valuable contributions, discussions, and support.

\appendix
\section{\label{sec:EHLagr} Formulation of the metric Lagrangian}
In this section, we derive the form of the Einstein--Hilbert action, $\sqrt{-g}R/2\kappa$, within the QR framework. In the theory, canonical quantization is applied to each quantum field, including the metric tensor. Meanwhile, $\delta \mathcal{S}/\delta g_{\mu \nu}$ yields $T^{\mu \nu}/2$, which served as an initial inspiration for promoting $g_{\mu \nu}$ to a dynamical field, inducing the QR condition even in the classical limit. Given that $T^{\mu \nu}$ is both the total stress--energy tensor and the sum of canonical momenta, this requirement introduces a unique constraint on the action. 

Throughout this work, we employ the vierbein formalism instead of the metric itself purely as a computational tool. The vierbein is assumed to be entirely determined by the metric, solely to avoid accounting for the symmetric nature of $g_{\mu \nu}$. Referring to $E_\mu ^i$ and $e^\mu _i$ as the vierbein and its inverse, respectively, and denoting by $E=\sqrt{-g}$ the determinant of $E_\mu ^i$, Eq.~\ref{eq:T} can be re-expressed as:
\begin{equation}
	(T^{(n)})^\mu _i = E e^\mu _i \Lagr_0 - \sum_n \frac{\partial (E \Lagr_0)}{\partial (\partial_\mu \phi_n)} \partial_i \phi_n \equiv E (e^\mu _i \Lagr_0 - N_i ^\mu)
\end{equation}
Here, the world indices and the Lorentz indices are distinguished using Greek and Latin alphabets, respectively. To describe the terms more conveniently, $\Lagr_0 \equiv \Lagr / E$ and $N_i ^\mu$ are newly defined. Then, the vierbein version of the variational derivative of the action leads to:
\begin{flalign}
	\frac{\delta \mathcal{S}}{\delta E^a _\alpha} &= E e^\alpha _a \Lagr_0 + E \frac{\delta}{\delta E^a _\alpha} \int d^4 x \Lagr_0 \label{eq:EHLagr} \\
	&= \sum_{n \neq g} T^{(n) \alpha} _a + T^{(g) \alpha} _a + e^\beta _a \partial_\gamma S^{\alpha \gamma} _{\ \ \ \beta}  \label{eq:EHLagr2}
\end{flalign} 
$\partial_\gamma S^{\alpha \gamma}_{\ \ \ \beta}$ is a total derivative term that serves as a boundary condition for $P^\alpha _\beta$, which disappears in the large-volume limit. Meanwhile, because of the well-known relation, $\delta \mathcal{S}_\text{QFT}/\delta E_\alpha ^a = \sum_{n \neq g} T^{(n) \alpha} _a$, Eq.~\ref{eq:EHLagr} must match $E e^\alpha _a \Lagr_0 - E \sum_{n \neq g} N^{(n) \alpha} _a$ when the spacetime is Minkowskian. By using this, and by defining $G_{\vec{p}}$ as the derivative-related terms for the metric, with $\vec{p}$ denoting a pair of indices, Eq.~\ref{eq:EHLagr} and Eq.~\ref{eq:EHLagr2} are equated as follows.
\begin{equation}
	\label{eq:EHequation}
	E \frac{\partial G_{\vec{p}}}{\partial E_\alpha ^a} \frac{\partial \Lagr_0}{\partial G_{\vec{p}}} - \partial_{\gamma} \Big( E \frac{\partial G_{\vec{p}}}{\partial (\partial_\gamma E_\alpha ^a)} \frac{\partial \Lagr_0}{\partial G_{\vec{p}}} \Big) = - E \frac{\partial G_{\vec{p}}}{\partial (\partial_\alpha E_\mu ^i)} \frac{\partial \Lagr_0}{\partial G_{\vec{p}}} \partial_a E_\mu ^i + e_a ^\beta \partial_\gamma S^{\alpha \gamma} _{\ \ \ \beta} 
\end{equation}
Choosing $S^{\alpha \gamma} _{\ \ \ \beta}$ like below:
\begin{equation}
	S^{\alpha \gamma} _{\ \ \ \beta} = - E E_\beta ^b \frac{\partial G_{\vec{p}}}{\partial (\partial_\gamma E_\alpha ^b)} \frac{\partial \Lagr_0}{\partial G_{\vec{p}}}
\end{equation}
Eq.~\ref{eq:EHequation} is solved if:
\begin{equation}
	\label{eq:Gcond}
	E_\beta ^a \frac{\partial G_{\vec{p}}}{\partial E_\alpha ^a} + \frac{\partial G_{\vec{p}}}{\partial (\partial_\gamma E_\alpha ^a)} \partial_\gamma E_\beta ^a + \frac{\partial G_{\vec{p}}}{\partial (\partial_\alpha E_\mu ^i)} \partial_\beta E_\mu ^i = 0
\end{equation}
The solution to the above equation is $G_{\vec{p}}=G_{ij} ^{\ \ k} = e^\mu _i e^\nu _j \partial_\mu E^k _\nu$. This equivalently indicates that curved spacetime can interact with other fields through either Christoffel symbols contracted to Lorentz indices, $\Gamma^{k} _{ij}$, or spin connections, $\omega_{k} ^{ij}$, as these provide another basis for representing $G_{ij} ^{\ \ k}$. Furthermore, since $G_{\vec{p}}$ is the variable describing $\partial_\gamma E_\alpha ^a$ on its behalf, the degrees of freedom match. As a result, $\Gamma^{k}_{ij}$ and $\omega_{k}^{ij}$ form a complete building block for the derivative terms.

Although there exists no universally governing Riemannian manifold, as discussed in Sec.~\ref{sec:2}, the coordinates for quantum fields are still described within a Riemannian framework, which serves as a tool for encoding the locality of the fields. It is then natural to require the equations of motion to be tensorial. Under this condition, and based on the given building blocks, the only metric terms (besides the $E$ and $E_\alpha ^{a}$ couplings involved in forming $\sum_{n \neq g} T^{(n)\alpha}_{a}$) that may be contained in the Lagrangian are the generalized covariant derivatives, $D_i$. Accordingly, the self-interacting terms for the metric must also be defined from $D_i$, and the straightforward approach to achieve this is to introduce the Riemann tensor, $R^{ij}_{kl}$. Now the simplest way to make a scalar is to contract the indices and obtain the Ricci scalar, $R = R^{ij}_{ij}$, recovering the Einstein--Hilbert Lagrangian. This implies that the requirement to fulfill the QR condition in the classical limit naturally leads to diffeomorphism invariance in each constructed Riemannian manifold.

\section{\label{sec:graviton} Rearrangement of the Lagrangian}
Starting from Eq.~\ref{eq:LgravStart}, the graviton Lagrangian can be rearranged into a more familiar form. By computing the derivatives:
\begin{flalign}
		-{2\kappa \Lagr_\lambda \over \sqrt{-g}} &= \bar{\lambda}_{\mu \nu} \frac{\delta}{\delta g_{\alpha \beta}} \big( G^{\mu \nu} + g^{\mu \nu} \Lambda \big)  \lambda_{\alpha \beta} + {1 \over 2} G^{\mu \nu} \bar{\lambda}_{\mu \nu} \lambda \\
	\label{eq:freeL}
	&\equiv \bar{\lambda}_{\mu \nu} X^{\alpha \beta \mu \nu} \lambda_{\alpha \beta} - {1 \over 2} \bar{\lambda} X^{\alpha \beta} \lambda_{\alpha \beta} + Y^{\alpha \beta \mu \nu} \bar{\lambda}_{\mu \nu} \lambda_{\alpha \beta}
\end{flalign}
Here, $\bar{\lambda}$ and $\lambda$ are the traces of the graviton field. Also, $X^{\alpha \beta \mu \nu} = \delta R^{\mu \nu} / \delta g_{\alpha \beta}$ and $X^{\alpha \beta} = X^{\alpha \beta \mu \nu} g_{\mu \nu}$ are newly defined for simpler expression. The rest terms are collected to define $Y^{\alpha \beta \mu \nu}$,
\begin{flalign}
		Y^{\alpha \beta \mu \nu}  \bar{\lambda}_{\mu \nu} \lambda_{\alpha \beta} &= \big({R \over 2} - \Lambda \big) \bar{\lambda}_{\mu \nu} \lambda^{\mu \nu} -{1 \over 2} R^{\alpha \beta} \bar{\lambda} \lambda_{\alpha \beta} + {1 \over 2} G^{\mu \nu} \bar{\lambda}_{\mu\nu} \lambda \\
	\label{eq:Y}
	&= \big({R \over 2}  - \Lambda \big) \big(\bar{\lambda}_{\mu \nu} \lambda^{\mu \nu} - {\bar{\lambda}\lambda \over 2} \big) + {R^{\alpha \beta} \over 2} (\bar{\lambda}_{\alpha \beta} \lambda -\bar{\lambda} \lambda_{\alpha \beta})
\end{flalign}
while $X^{\alpha \beta \mu \nu}$ terms are explicitly expressed as follows.
\begin{flalign}
		(X^{\alpha \beta \mu \nu} \ \mathrm{terms}) &= \big(\bar{\lambda}_{\mu \nu} - {1 \over 2} \bar{\lambda} g_{\mu \nu} \big) \frac{\delta}{\delta g_{\alpha \beta}} (g^{\mu \rho} g^{\nu \sigma} R_{\rho \sigma}) \lambda_{\alpha \beta} \\
	\label{eq:X}
	&= R^{\alpha \beta} (\bar{\lambda} \lambda_{\alpha \beta }- 2 \bar{\lambda}_{\mu \alpha} \lambda^\mu _\beta ) + \big( \bar{\lambda}^{\mu \nu} - {\bar{\lambda} g^{\mu \nu} \over 2} \big) \frac{\delta R_{\mu \nu}}{\delta g_{\alpha \beta}} \lambda_{\alpha \beta}
\end{flalign}
To find $(\delta R_{\mu \nu}/\delta g_{\alpha \beta})\lambda_{\alpha \beta}$, $(\delta \Gamma^\gamma _{\mu \nu} / \delta g_{\alpha \beta}) \lambda_{\alpha \beta}$ is firstly explored as a building block ($\Gamma^\gamma _{\mu \nu}$ are the Christoffel symbols).
\begin{flalign}
		\frac{\delta \Gamma^\gamma _{\mu \nu}}{\delta g_{\alpha \beta}} \lambda_{\alpha \beta} &= \frac{\delta}{\delta g_{\alpha \beta}} \big( {g^{\gamma \delta} \over 2} (\partial_\mu g_{\nu \delta} + \partial_\nu g_{\mu \delta} - \partial_\delta g_{\mu \nu}) \big) \lambda_{\alpha \beta} \\
	&= -\lambda^\gamma _\delta \Gamma^\delta _{\mu \nu} + {g^{\gamma \delta} \over 2} (\partial_\mu \lambda_{\nu \delta} + \partial_\nu \lambda_{\mu \delta} - \partial_\delta \lambda_{\mu \nu}) \\
	&= {1 \over 2} (\nabla_\mu \lambda_\nu ^\gamma + \nabla_\nu \lambda_\mu ^\gamma - \nabla^\gamma \lambda_{\mu \nu})
\end{flalign}
Using the above it becomes obvious that:
\begin{equation}
	\begin{split}
		2\frac{\delta}{\delta g_{\alpha \beta}} (\partial_\gamma \Gamma^\gamma _{\mu \nu}) \lambda_{\alpha \beta} &= \nabla_\gamma (\nabla_\mu \lambda_\nu ^\gamma + \nabla_\nu \lambda_\mu ^\gamma - \nabla^\gamma \lambda_{\mu \nu})\\
		& +(\Gamma^\delta _{\gamma \mu} \nabla_\delta \lambda^\gamma _\nu + \Gamma^\delta _{\gamma \nu} \nabla_\mu \lambda ^\gamma _\delta - \Gamma^\gamma _{\gamma \delta} \nabla_\mu \lambda^\delta _\nu ) + (\mu \leftrightarrow \nu) \\
		& + (\Gamma ^\gamma _{\gamma \delta} \nabla ^\delta \lambda_{\mu \nu} - \Gamma^\delta _{\gamma \mu} \nabla^\gamma \lambda_{\delta \nu} - \Gamma^\delta _{\gamma \nu} \nabla^\gamma \lambda_{\delta \mu})
	\end{split}
\end{equation}
and,
\begin{equation}
	-2\frac{\delta}{\delta g_{\alpha \beta}} (\partial_\nu \Gamma^\gamma _{\gamma \mu}) \lambda_{\alpha \beta} = -\nabla_\nu \nabla_\mu \lambda - \Gamma^\gamma _{\mu \nu} \nabla_\gamma \lambda
\end{equation}
and also,
\begin{equation}
	\begin{split}
		2 \frac{\delta}{\delta g_{\alpha \beta}} (\Gamma^\gamma _{\gamma \delta} \Gamma^\delta _{\mu \nu} - \Gamma^\delta _{\mu \gamma} \Gamma^\gamma _{\delta \nu}) \lambda_{\alpha \beta}	&= \Gamma^\gamma _{\mu \nu} \nabla_\gamma \lambda + \Gamma^\gamma _{\gamma \delta} (\nabla_\mu \lambda ^\delta _\nu + \nabla_\nu \lambda^\delta _\mu - \nabla^\delta \lambda_{\mu \nu}) \\
		&+\big(- \Gamma^\delta _{\mu \gamma}(\nabla_\delta \lambda^\gamma _\nu + \nabla_\nu \lambda^\gamma _\delta - \nabla^\gamma \lambda_{\nu \delta}) \big) + (\mu \leftrightarrow \nu)
	\end{split}
\end{equation}
Combining them all, one finds the desired form.
\begin{flalign}
	\begin{split}
		\frac{\delta R_{\mu \nu}}{\delta g_{\alpha \beta}} \lambda_{\alpha \beta} &= \frac{\delta}{\delta g_{\alpha \beta}} (\partial_\gamma \Gamma^\gamma _{\mu \nu} - \partial_\nu \Gamma^\gamma _{\gamma \mu} + \Gamma^\gamma _{\gamma \delta} \Gamma^\delta _{\mu \nu} - \Gamma^\delta _{\mu \gamma} \Gamma^\gamma _{\delta \nu}) \lambda_{\alpha \beta}
	\end{split} \\
	\label{eq:dRdg}
	&= {1 \over 2} (\nabla_\gamma \nabla_\mu \lambda^\gamma _\nu + \nabla_\gamma \nabla_\nu \lambda^\gamma _\mu - \nabla_\gamma \nabla^\gamma \lambda_{\mu \nu} - \nabla_\nu \nabla_\mu \lambda)
\end{flalign}
From Eq.~\ref{eq:dRdg}, the exact form for the kinetic terms in Eq.~\ref{eq:X} is found.
\begin{equation}
	\label{eq:GravKinetic}
	\begin{split}
		\text{(Ki} \text{netic terms)} &= {1 \over 2} \nabla^\gamma \bar{\lambda}^{\mu \nu} \nabla_\gamma \lambda_{\mu \nu} - \nabla^\gamma \bar{\lambda}^{\mu \nu} \nabla_\mu \lambda_{\gamma \nu} \\
		&+ {1 \over 2} \nabla_\nu \bar{\lambda}^{\mu \nu} \nabla_\mu \lambda + {1 \over 2} \nabla_\nu \bar{\lambda} \nabla_\mu \lambda^{\mu \nu} - {1 \over 2} \nabla_\gamma \bar{\lambda} \nabla^\gamma \lambda
	\end{split}
\end{equation}
The second term can be rewritten as follows (vanishing total derivatives are used).
\begin{flalign}
	\begin{split}
		- \nabla^\gamma \bar{\lambda}^{\mu \nu} \nabla_\mu \lambda_{\gamma \nu} &= (\nabla_\mu \nabla_\gamma \bar{\lambda}^{\mu \nu}) \lambda_\nu ^\gamma
	\end{split} \\
	&= (\nabla_\gamma \nabla_\mu \bar{\lambda}^{\mu \nu}) \lambda^\gamma _\nu + (\nabla_{[\mu} \nabla_{\gamma]} \bar{\lambda}^{\mu \nu}) \lambda^\gamma _\nu \\
	\label{eq:lambdaHalf}
	&= -\nabla_\mu \bar{\lambda}^{\mu \nu} \nabla_\gamma \lambda^\gamma _\nu + R_{\mu \nu} \bar{\lambda}^{\mu \gamma} \lambda^\nu _\gamma + R^\nu _{\ \alpha \mu \gamma} \bar{\lambda}^{\alpha \mu} \lambda^\gamma _\nu
\end{flalign}
By applying Eq.~\ref{eq:lambdaHalf} to the half of $\nabla^\gamma \bar{\lambda}^{\mu \nu} \nabla_\mu \lambda_{\gamma \nu}$ in Eq.~\ref{eq:GravKinetic}, the kinetic terms can be expressed with the more symmetric manner. Together with Eqs.~\ref{eq:freeL}, \ref{eq:Y}, \ref{eq:X}, and \ref{eq:dRdg}, the full expression for the graviton's free Lagrangian, $\Lagr_{\lambda}$ in Eq.~\ref{eq:Lgrav}, is derived. According to our calculations, this result remains unchanged even when the vierbein formalism is employed, provided that the theory is kept fully metric; only the overall factor changes from $1/4$ to $1$.

\section{\label{sec:propagator} Free graviton propagators}
Here, we compute and obtain the graviton propagators based on Eq.~\ref{eq:propStart}. To begin with, the Grassmanian variable $\lambda_{ij}$ in $Z_\lambda$ is integrated using $\int D\lambda D\bar{\lambda} \exp(\bar{\lambda} \cdot A \cdot \lambda) = \det|A|$. After normalizing the partition function,
\begin{equation}
	Z_\lambda = \int D\chi D\bar{\chi} e^{i \int d^4 p \Lagr_\chi(p)}
\end{equation}
remains with the leftover Lagrangian, $\Lagr_\chi(p)$.
\begin{equation}
	\Lagr_\chi(p) = (\bar{J}^{(ij)} + \bar{\chi}^a P^{(ij)}_a)G_{ijkl}(J^{(kl)} + P^{(kl)} _b \chi^b)
\end{equation}
$G_{ijkl}$ indicates the inverse of $A^{ijkl}$. For easier calculation, it is better to change the basis of $A$: $A^{ijkl} \rightarrow A^{(ij)(kl)}$.
\begin{flalign}
	{4 \over p^2}A^{(ij)(kl)} = \begin{pNiceMatrix}[first-row,first-col,code-for-first-row=\scriptstyle, code-for-first-col=\scriptstyle]
		& (01) & (02) & (03) & (12) & (13) & (23) \\
		(01) & -2 & 0 & 0 & 0 & 0 & 0 \\
		(02) & 0 & -2& 0 & 0 & 0 & 0 \\
		(03) & 0 & 0 & -2 & 0 & 0 & 0 \\
		(12) & 0 & 0 & 0 & 2 & 0 & 0 \\
		(13) & 0 & 0 & 0 & 0 & 2 & 0 \\
		(23) & 0 & 0 & 0 & 0 & 0 & 2 
	\end{pNiceMatrix}
\end{flalign}
and,
\begin{flalign}
	{4 \over p^2}A^{(ii)(kk)} = \begin{pNiceMatrix}[first-row,first-col,code-for-first-row=\scriptstyle, code-for-first-col=\scriptstyle]
		& (00) & (11) & (22) & (33) \\
		(00) & 0 & 1 & 1 & 1 \\
		(11) & 1 & 0 & -1 & -1 \\
		(22) & 1 & -1 & 0 & -1 \\
		(33) & 1 & -1 & -1 & 0 
	\end{pNiceMatrix}
\end{flalign}
Other terms are zeros. The inverse of $A^{(ij)(kl)}$, $G_{(ij)(kl)}$, is easily computed using block matrices.
\begin{equation}
	G_{(ij)(kl)} = {2 \over p^2} 
	\begin{pmatrix} 
		\begin{pmatrix} -I_3 &  \\ & I_3 \end{pmatrix} &   \\
		& {2 \over 3} \begin{pmatrix} 2 & 1 & 1& 1 \\ 1 & 2 & -1 & -1 \\ 1 & -1 & 2 & -1 \\ 1 & -1 & -1 & 2 \end{pmatrix}
	\end{pmatrix} 
\end{equation}
The empty places are zeros, and $I_3$ indicates $3 \times 3$ identity matrix. Now,
\begin{equation}
	\label{eq:chiInt}
	\begin{split}
		(\bar{J}^{(ij)} + \bar{\chi}^a P^{(ij)}_a)G_{ijkl}(J^{(kl)} + P^{(kl)} _b \chi^b) &= \bar{J}^{(ij)} G_{(ij)(kl)} J^{(kl)} \\
		&\quad - \bar{J}^{(ij)} G_{(ij)(kl)} P^{(kl)} _a C^{ab} P^{(wx)} _b G_{(wx)(yz)} J^{(yz)} \\
		&\quad + \bar{D}^a B_{ab} D^b
	\end{split}
\end{equation}
where
\begin{flalign}
	B_{ab} &= P^{(ij)}_a G_{(ij)(kl)} P^{(kl)}_b \\
	\label{eq:Bmatrix}
	&= {2 \over 3p^2} \begin{pmatrix}
		4p_0 ^2 - 3 |\vec{p}|^2 & p_0 \cdot p_1 & p_0 \cdot p_2 & p_0 \cdot p_3 \\
		p_0 \cdot p_1 & p_{231} & p_1 \cdot p_2 & p_1 \cdot p_3 \\
		p_0 \cdot p_2 & p_1 \cdot p_2 & p_{132} & p_2 \cdot p_3 \\
		p_0 \cdot p_3 & p_1 \cdot p_3 & p_2 \cdot p_3 & p_{123}
	\end{pmatrix}
\end{flalign}
and $C = B^{-1}$, with
\begin{equation}
	D^b = \chi^b + C^{bc} P^{(ij)} _c G_{(ij)(kl)} J^{(kl)}
\end{equation}
Here, $p_{ijk} \equiv 3(p_i ^2 + p_j ^2 - p_0 ^2) + 4p_k ^2$ is defined and used. Taking the inverse of Eq.~\ref{eq:Bmatrix}, $C^{ab}$ is also obtained.
\begin{equation}
	C^{ab} = {9 \over 2} \eta^{ab} + {3 \over 2p^2} |p^a p^b|
\end{equation}
Using the fact that the last term in Eq.~\ref{eq:chiInt} is a constant after $\chi$ integration, the normalized $Z_\lambda$ becomes:
\begin{equation}
	Z_\lambda = \exp \Big(i \int d^4 p \bar{J}^{(ij)} \Delta_{(ij)(xy)} J^{(xy)} \Big)
\end{equation}
where $\Delta_{(ij)(xy)}$ are the graviton propagators.
\begin{flalign}
	\Delta &\equiv G_{(ij)(kl)} \big(I^{(kl)} _{(xy)} - P^{(kl)} _a C^{ab} P^{(uv)} _b G_{(uv)(xy)}\big) \\
	&= {2 \over p^6} \begin{pNiceMatrix}[first-row,last-col=3,code-for-first-row=\scriptstyle, code-for-last-col=\scriptstyle] (x \neq y) &  (x=y) &  \\ \Delta^1 & \Delta^3 & (i \neq j) \\ (\Delta^3)^\dagger & \Delta^2 & (i = j) \end{pNiceMatrix} 
\end{flalign}
The values for $\Delta^1$, $\Delta^2$, and $\Delta^3$ are:
\begin{equation}
	\Delta^1 _{(ij)(xy)} = 
	\begin{cases}
		\eta^{ii} \eta^{jj} (p^2 - \eta^{ii}p_i ^2)(p^2 - \eta^{jj} p_j ^2)&  (ij)(ij)\\
		\eta^{ii} p_j p_k (\eta^{ii} p_i ^2 - p^2)&  (ij)(ik), \ j \neq k \\
		p_0 p_1 p_2 p_3& \text{else}
	\end{cases}
\end{equation}
and,
\begin{equation}
	\Delta^2 _{(ii)(jj)} = 
	\begin{cases}
		(p^2 - \eta^{ii}p_i ^2)^2 & i=j \\
		p_i ^2 p_j ^2 - \eta^{ii} \eta^{jj} p^2 (p^2 - \eta^{ii} p_i ^2 - \eta^{jj} p_j ^2) & \text{else}
	\end{cases}
\end{equation}
with,
\begin{equation}
	\Delta^3 _{(ij)(kk)} = 
	\begin{cases}
		\eta^{kk} p_i p_j (\eta^{kk} p_k ^2 - p^2) & i=k \text{ or } j=k \\
		\eta^{kk} p_i p_j (\eta^{kk}p_k ^2 + p^2) & \text{else}
	\end{cases}
\end{equation}
These terms can also be expressed in another way, by dividing $\Delta_{(ij)(xy)}$ into 5 categories, which is represented in Eq.~\ref{eq:prop}.

\section{\label{sec:propPosition} Fourier transform of $1/p^6$}

This section provides the Fourier transform of $1/p^6$, which serves as a basis for identifying and generating causal graviton propagators. Because there is anti-$t$ symmetry for causal propagators, let's assume $t>0$ first and find $\Delta(x)$ without loss of generality.

\begin{equation}
	\FT^{-1}\{ {(2\pi)^4 \over p^6} \} = \int d^3 \vec{p} \oint dp_0 {\exp(ip \cdot x) \over (p_r ^2 - p_0 ^2)^3}
\end{equation}
$\oint dp_0$ denotes the counterclockwise closed loop integration over the complex plane. Using the residue theorem, the above equation is computable.
\begin{flalign}
	\FT^{-1}\{ {(2\pi)^4 \over p^6} \} &= {1 \over 2} \cdot 2\pi i \int d^3 \vec{p} {e^{i \vec{p} \cdot \vec{x}} \over (2p_r)^3} \Big(it^2 \sin(p_r t) -3i {\sin(p_r t) \over p_r ^2} + 3it {\cos(p_r t) \over p_r} \big) \\
	&= {\pi^2 \over 4} \int_0 ^\infty dp_r \int_{-1} ^1 d \cos \theta e^{ip_r r \cos \theta} \Big(\big({3 \over p_r ^3} - {t^2 \over p_r}\big) \sin (p_r t) - 3t {\cos (p_r t) \over p_r ^2} \Big)
\end{flalign}
After the integration of $\theta$:
\begin{equation}
	\label{eq:Fp6}
	\FT^{-1}\{ {(2\pi)^4 \over p^6} \} = {\pi^2 \over 2r} \int_0 ^\infty dp_r \Big( \big({3 \over p_r ^4} - {t^2 \over p_r ^2}\big) \sin (p_r t) \sin(p_r r)  - {3t \over p_r ^3} \cos (p_r t) \sin(p_r r) \Big)
\end{equation}
Each integral in the above diverges, but the integration of the total function does not. To handle this, a small $\varepsilon$ is introduced. Using integral by parts to the first term,
\begin{equation}
	\int_\varepsilon ^\infty dq {\sin(qt) \sin(qr) \over q^4} = \Big[ -{\sin(qt) \sin(qr) \over 3q^3}  \Big]_\varepsilon ^\infty + \int_\varepsilon ^\infty dq {t \cos (qt) \sin(qr) + r \sin (qt) \cos (qr) \over 3q^3}
\end{equation}
Repeating integral by parts two times more:
\begin{equation}
	\label{eq:varEpTrick}
	\begin{split}
		\text{F} \text{irst term} = {tr \over \varepsilon} &- {t^2 + r^2 \over 6} \int_\varepsilon ^\infty dq {\sin(qt) \over q} {\sin(qr) \over q} \\
		&- {tr \over 3} \int_\varepsilon ^\infty {dq \over q} (t \sin (qt) \cos (qr) + r \cos (qt) \sin(qr))
	\end{split}
\end{equation}
The result of the first integral in Eq.~\ref{eq:varEpTrick} is,
\begin{equation}
	\lim_{\varepsilon \to 0} \int_\varepsilon ^\infty dq {\sin(qt) \over q} {\sin(qr) \over q} = {\pi \over 4} (|r+t| - |r-t|)
\end{equation}
Similarly, the limit after the second integral is,
\begin{equation}
	{\pi \over 4} (t \sgn(r+t) - t \sgn(r-t)+ r \sgn(r+t) - r \sgn(t-r))
\end{equation}
$\sgn(x)$ is a signum function, giving 1 if $x>0$, -1 if $x<0$, and 0 if $x=0$. Now substituting the results to Eq.~\ref{eq:varEpTrick} with the multiplied factors from Eq.~\ref{eq:Fp6},
\begin{equation}
	\text{First term} = {3\pi^2 t \over 2\varepsilon} - {\pi^3 \over 8}
	\begin{cases}
		(t^2 + 3r^2) \cdot t/r &r > t\\
		3t^2 + r^2 & r \le t
	\end{cases}
\end{equation}
By applying the aforementioned integrals, both the second and third terms in Eq.~\ref{eq:Fp6} can be calculated as well.
\begin{equation}
	\text{Second term} = - \pi^3 {t^2 \over 8r} (|r+t| - |r-t|)
\end{equation}
\begin{equation}
	\text{Third term} = -{3\pi^2 t \over 2 \varepsilon} + {3\pi^3 \over 8}
	\begin{cases}
		t^3/r + tr & r > t \\
		2t^2 & r \le t
	\end{cases}
\end{equation}
By adding three terms, Eq.~\ref{eq:Fp6} is found to converge, and the transformation is completed after restoring the anti-$t$ symmetry (see Eq.~\ref{eq:Fp6Results}).

\bibliography{QR}

\end{document}